\documentclass[iop,apj]{emulateapj}
\usepackage{amsmath }
\usepackage{epstopdf}

\shortauthors{Ruan, Anderson, Plotkin et al.}
\shorttitle{The Nature of Transition Blazars} 
\begin{document}
\title{The Nature of Transition Blazars}
\author{J~J.~Ruan\altaffilmark{1,2}, 
  S.~F.~Anderson\altaffilmark{2}, 
  R.~M.~Plotkin\altaffilmark{3}, 
  W.~N.~Brandt\altaffilmark{4,5}, 
  T.~H.~Burnett\altaffilmark{6},
  A.~D.~Myers\altaffilmark{7},
  D.~P.~Schneider\altaffilmark{4,5}
  }
\altaffiltext{1}{ Corresponding author: jruan@astro.washington.edu} 
\altaffiltext{2}{Department of Astronomy, University of
Washington, Box 351580, Seattle, WA 98195, USA}
\altaffiltext{3}{Department of Astronomy, University of Michigan, 500
Church St., Ann Arbor, MI 48109, USA}
\altaffiltext{4}{Department of Astronomy \& Astrophysics, 525 Davey Lab, The 
Pennsylvania State University, University Park, PA 16802, USA}
\altaffiltext{5}{Institute for Gravitation and the Cosmos, The Pennsylvania State 
University, University Park, PA 16802, USA}
\altaffiltext{6}{Department of Physics, University of Washington, Seattle, WA 98195-1560, USA}
\altaffiltext{7}{University of Wyoming, Dept. of Physics and Astronomy 3905, 1000 E. University, Laramie, WY 82071, USA}
\keywords{blazars: general}

\begin{abstract}
	Blazars are classically divided into the BL Lac (BLL) and Flat-Spectrum Radio Quasar 
(FSRQ) subclasses, corresponding to radiatively inefficient and efficient accretion regimes, respectively,
largely based on the equivalent width (EW) of their optical broad emission lines (BEL). However, 
EW-based classification criteria are not physically motivated, and a few blazars have previously 
`transitioned' from one subclass to the other. We present the first systematic search for these 
transition blazars in a sample of 602 unique pairs of repeat spectra of 354 blazars in SDSS, finding six 
clear cases. These transition blazars have bolometric Eddington ratios of $\sim$0.3 and low-frequency 
synchrotron peaks, and are thus FSRQ-like. We show that the strong EW variability (up to an 
unprecedented factor of $>$60) is due to swamping of the BELs from variability in jet continuum 
emission, which is stronger in amplitude and shorter in timescale than typical blazars. 
Although these transition blazars appear to switch between FSRQ and BLL according to the 
phenomenologically-based EW scheme, we show that they are most likely rare cases of FSRQs 
with radiatively efficient accretion flows and especially strongly-beamed jets.  These results have 
implications for the decrease of the apparent BLL population at high-redshifts, and may add credence
 to claims of a negative BLL redshift evolution.
\end{abstract}

\section{Introduction}

	Blazars are a relatively rare class of Active Galactic Nuclei (AGN) in which a jet is aligned 
closely along the observed line of sight \citep{bl78}. These jets are relativistically beamed, leading 
to unique multi-wavelength properties from the radio through the $\gamma$-rays \citep{an93, ur95}. 
Classically, blazars have been divided into the Flat-Spectrum Radio Quasar (FSRQ) and BL Lac (BLL) 
subclasses based on whether any optical broad emission lines (BELs) have rest-frame equivalent widths (EWs) 
above (in the case of FSRQs) or below (in the case of BLLs) 5\AA~\citep[e.g.][]{sti91, 
sto91, pe95, pl10}. These two subclasses of blazars also show systematic differences in 
Eddington ratios \citep{da07}, $\gamma$-ray properties \citep{ac11}, and the luminosities 
and frequencies of their synchrotron and inverse-Compton peaks of their spectral energy distributions 
\citep{pa95, sa96, fo98, ab10a}, amongst many other differences.

	The physical distinction between FSRQs and BLLs is suspected to stem from the divergent natures of their 
underlying accretion flows around the central supermassive black holes \citep{ma02, gh08}. FSRQs are believed 
to harbor geometrically-thin, optically-thick accretion disks that are accreting rapidly, with Eddington ratios 
$\gtrsim$0.1, similar to (and perhaps systematically higher than) typical Type 1 quasars \citep{sh12}.
The existence of broad emission lines in their optical spectra from high-velocity gas clouds, as well as strong emission 
in the infrared from dusty torii, are well-described by the standard radio-loud unification scheme of AGNs \citep{ur95}. In 
contrast, BLLs are thought to have geometrically-thick, optically-thin accretion disks (at least in their inner 
regions), where the accretion is radiatively inefficient \citep{na94}, with Eddington ratios generally $\lesssim$0.01 \citep{wa02, da07}. Their lack of broad emission lines and dust emission suggests that the properties of 
accretion flow and structure of the surrounding gas in BLLs are physically different from FSRQs.

	The classical EW-based classification of blazars between FSRQs and BLLs is not physically motivated, 
and a handful of blazars have been previously reported to transition between these two classes in repeat 
spectroscopic observations due to variability in their BEL EWs \citep[e.g.,][]{gh11, sh12}. These rare 
`transition' blazars ironically include the namesake BL Lacertae object, in which the rest-frame H$\alpha$ EW has 
been reported to vary between approximately 0 to $\sim$7\AA~\citep{ve95, co96}. Analysis of long-term 
spectroscopic monitoring of BL Lacertae by \citet{co00} demonstrated that this variability in the BEL EW is 
anti-correlated with the continuum luminosity, suggesting that the strongly-variable jet continuum emission 
is swamping the BEL flux, leading to the transition blazar phenomena. More recent reports of 
\emph{Fermi}-detected transition blazars by \citet{sh12} have provided observational motivation for 
investigations into the mysterious nature of these objects \citep[see also][]{is13}. \citet{gh11} showed that many 
\emph{Fermi}-detected BLLs with $\gamma$-ray properties similar to FSRQs also have SEDs and 
BEL luminosities that are FSRQ-like, thus suggesting that some BL Lacs may actually be FSRQs 
with particularly strong jet continuum emission. Based on their findings, \citet{gh11} also suggested an 
improved classification scheme based on the Eddington ratio calculated from the BEL luminosities; 
this approach is more physically-motivated and particularly helpful for transition blazars 
\citep[also see][]{gi12, sa12}.

	Correct classification of transition blazars is critical not only for studies of the different properties 
of FSRQs and BLLs \citep[e.g., along the `blazar sequence',][]{fo98}, but also has important implications 
for their redshift evolution. For example, BLLs are more commonly found at lower redshifts, leading 
to many reports of apparent negative redshift evolution \citep{ma84, wo94, re00, be03}, although this interpretation
has been often disputed \citep{ca02, pa07, ma13}. Misclassifications of blazars can artificially lead to 
an apparent lack of high-redshift BLLs, significantly affecting evolutionary scenarios of the blazar 
sequence \citep[e.g.,][]{bo02} and the blazar contributions to the cosmic $\gamma$-ray background \citep[e.g.,][]{ab10b}. 
These issues of correct classification face the further challenge of obtaining correct redshifts from the 
often featureless spectra of blazars; strong BEL EW variability in transition blazars can be used to 
determine or constrain redshifts in repeat spectra during epochs in which the EWs are large \citep{sh13b}.

	In this paper, we present the first systematic search for transition blazars in a large sample of 
repeat spectra of known blazars in the Sloan Digital Sky Survey \citep[SDSS,][]{yo00}. 
Many previous reports of transition blazars have been based on comparison to EW classifications in the 
literature; the homogeneous nature of SDSS spectra will allow a more robust measurement of their variable 
spectral properties over timescales of over a decade in the observed frame. To complement the repeat optical 
spectra in our study, we also adopt a multi-wavelength approach to investigate the $\gamma$-ray, optical 
variability, and SED properties of these rare objects, tying our findings to the many recent investigations of 
blazar properties in these domains. 

	The outline of this paper is as follows. In Section 2, we present our sample of repeat blazar 
spectra and spectral analysis to identify transition blazars. In Section 3, we discuss the spectral and 
multi-wavelength properties of the transition blazars in our sample. In Section 4, we comment on the 
individual transition blazars in our sample. In Section 5, we discuss the nature of these transition 
blazars based on their properties. We summarize and conclude in Section 6. Throughout
the paper, EW refers to the equivalent width calculated in the rest-frame. We assume a standard 
$\Lambda$CDM cosmology with $\Omega_M = 0.3$, $\Omega_\Lambda = 0.7$, and 
$H_0 = 70$ km s$^{-1}$ Mpc$^{-1}$, consistent with the WMAP 9-year results of \citet{hi13}.

\begin{figure}
\begin{center}
\includegraphics[width=0.47\textwidth]{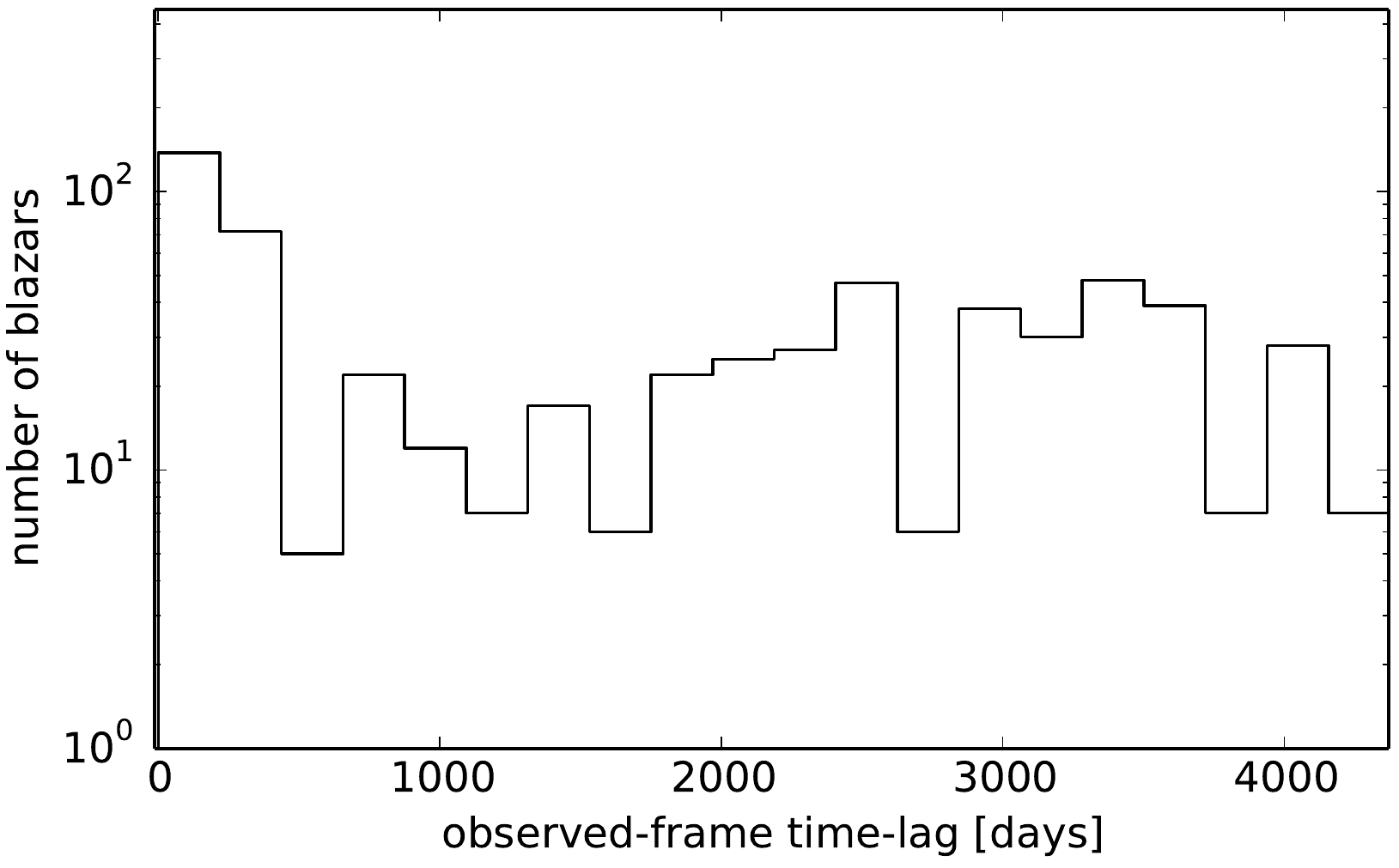} 
\caption{Histogram of the observed-frame time-lags between each pair of repeat
blazar spectra in our sample of 602 unique pairs. The median redshift of blazars
with measured redshifts is 1.05, although this value is strongly biased towards FSRQs, which 
have systematically higher redshifts than BL Lacs.}
\end{center}
\end{figure}

\section{Data and Search for Transition Blazars}
\subsection{Analysis of SDSS Spectra}

	Our data are taken from the SDSS-III \citep{gu06, ei11} Data Release 10 \citep[DR10,][]{ah13}, 
which is publicly available and also includes all data from the SDSS-I/II portions of the survey. DR10 
includes $\sim$4$\times$10$^6$ optical spectra over approximately 14,000 deg$^2$ of sky, including 
$\sim$1.6$\times$10$^5$ quasars \citep{pa14}, selected primarily from the imaging portion of the 
survey in a variety of science programs \citep{ro12}. In our investigation, we include all optical spectra taken 
from the primary survey portion of SDSS-I/II on the SDSS spectrograph, as well as spectra from the BOSS 
spectrograph used in SDSS-III (spectral resolution $R\sim2000$). Due to differences between the spectrographs,
fiber system, and spectral reduction pipeline between these phases of the SDSS survey 
\citep[see][]{bo12,sm13}, some small differences in absolute flux calibration may be present in spectra 
taken at different epochs \citep{da12}. These differences do not strongly affect our results, which are primarily 
based on spectral properties which are normalized to the continuum (e.g. equivalent widths). 

	Occasionally, repeat spectra of the same objects in SDSS were taken, either due to low signal-to-noise 
ratios in the first epoch, or as part of the survey plan. To create a sample 
of known blazars with repeat spectra in SDSS-III, we match all objects with repeat spectra available in DR10 
to the 3,149 blazars in the ROMA-BZCAT\footnote{http://www.asdc.asi.it/bzcat/} catalog of \citet{ma09}. 
This comparison results in a sample of 354 blazars with repeat observations, including a few with more than two 
epochs. Since the classical EW-based classification is based on whether any BELs in the observed optical spectrum 
are above or below 5\AA, we will compare the BEL EWs in pairs of repeat spectra of each blazar to identify 
cases where the BEL crossed this EW threshold. For blazars with more than two epochs of spectra, we will compare all 
unique pairs, leading to a total of 602 unique pairs of repeat spectra of the 354 blazars.
Blazars are known to be highly variable even in intra-night observations, and the timescales on which
blazars can transition between the FSRQ and BLL subclasses is not known; we thus place 
no cuts on the time-lag between repeat observations for our sample, which range from intra-night to 
approximately 12.5 years in the observed frame. The distribution of time-lags for our sample is shown
in Figure 1.

	We correct all repeat blazar spectra in our sample for Galactic extinction, using the
maps of \citet{sc98} and the Milky Way reddening law of \citet{ca89}. We also shift the repeat spectra of each 
blazar into the rest-frame using redshifts provided by the ROMA-BZCAT catalog, cognizant of the fact that the 
redshifts of many blazars in this catalog are missing, uncertain, or occasionally incorrect. To ensure that these 
redshifts we use are not obviously incorrect, we visually inspect all 602 unique pairs of repeat blazar spectra in 
our sample. In the visual inspection, we mask pixels in each spectrum with SDSS flags set for NOPLUG, 
BADTRACE, BADFLAG, BADARC, MANYBADCOLUMNS, MANYREJECTED, NEARBADPIXEL, LOWFLAT, 
FULLREJECT, SCATTEREDLIGHT, NOSKY, BRIGHTSKY, COMBINEREJ, REDMONSTER 
\citep[for details on these flags, see][]{st02}. Due to this spectrum quality mask, 16 of the 602 unique pairs
of repeat blazar spectra in our sample were removed.

	As expected, the visual inspection reveals that the overwhelming majority of blazars appear to have 
spectra that are either featureless (canonical BLLs) or have moderately-strong BELs (canonical FSRQs). 
Some outliers include BLLs that possessed strong host-galaxy contributions (e.g., misaligned BLLs), blazars 
with intrinsic and line-of-sight absorption features, as well as a small handful in which the emission lines 
strongly varied in EW between the different spectral epochs. To quantitatively identify blazars which have 
transitioned from FSRQs to BLLs (or vice versa) in our sample, we measure the properties of the 
C~IV $\lambda$1546, C~III] $\lambda$1906, Mg~II $\lambda$2800, and H$\alpha$ $\lambda$6565 broad 
emission lines (in vacuums wavelengths), where possible. 	

\begin{deluxetable*}{cccccccccccc}
\tablecolumns{12}
\tablewidth{0pt}
\tablecaption{SDSS spectral properties of all broad emission lines present in each of the transition blazars in our sample.}
\tablehead {\colhead{SDSS} & \colhead{$z$} & \colhead{Emission} & \colhead{MJD} & \colhead{EW} & \colhead{log$_{10}$$L_{\rm{line}}$} & \colhead{FWHM} & \colhead{log$_{10}$$M_{\rm{BH}}$} & \colhead{log$_{10}$($L_{\rm{bol}}$} & \colhead{log$_{10}$($L_{\rm{BLR}}$} \\ 
\colhead{Object} & & \colhead{Line} & & \colhead{[\AA]} & \colhead{[erg s$^{-1}$]} & [km s$^{-1}$] & \colhead{[M$_{\odot}$]} & \colhead{/$L_{\rm{Edd}})$} & \colhead{/$L_{\rm{Edd}})$}
}
\startdata
J083223.22+491321.0$^\textrm{a}$ & 0.548$^\textrm{b}$ & Mg~II & 51873 & 0.9 $\pm$ 0.7 & 41.52 $\pm$ 0.02 & & & & \\
& & & 55277 & 1.3 $\pm$ 0.3 & 41.47 $\pm$ 0.01 & & & & \\
& & & 55290 & 2.3 $\pm$ 0.5 & 41.59 $\pm$ 0.01 & 2126 $\pm$ 332 & 6.84 $\pm$ 0.15 & -0.76 $\pm$ 0.23 &  -2.14 $\pm$ 0.15 \\
\rule{0pt}{2ex}  

J083353.88+422401.8 & 0.249$^\textrm{c}$ & H$\alpha$ & 52266 & 5.1 $\pm$ 0.6 & 41.63 $\pm$ 0.05 & 809 $\pm$ 228 & 6.33 $\pm$ 0.26 & 0.18 $\pm$ 0.48 & -1.94 $\pm$ 0.26 \\ 
& & & 54524 & 4.8 $\pm$ 0.8 & 41.71 $\pm$ 0.07 & & & & \\
& & & 55924 & 2.1 $\pm$ 0.3 & 41.68 $\pm$ 0.06 & & & & \\
\rule{0pt}{2ex}  

J101603.13+051302.3 & 1.714$^\textrm{c}$ & C~IV & 52347 & 1.8 $\pm$ 0.2 & 43.12 $\pm$ 0.04 & & & \\
& & & 52355 & 14.8 $\pm$ 0.4 & 43.91 $\pm$ 0.01 & & & & \\
& & & 52356 & 43.4 $\pm$ 1.6 & 43.94 $\pm$ 0.01 & & & & \\
& & & 52366 & 19.56 $\pm$ 5.3 & 43.70 $\pm$ 0.09 & & & & \\
& & & 55653 & 116.8 $\pm$ 4.3 & 43.76 $\pm$ 0.01 & 3267 $\pm$ 50 & 8.44 $\pm$ 0.22 &  -0.54 $\pm$ 0.61 & -1.83 $\pm$ 0.22 \\
\rule{0pt}{2ex}  

& & C~III] & 52347 & 1.0 $\pm$ 0.1 & 42.75 $\pm$ 0.05 & & & & \\
& & & 52355 & 4.0 $\pm$ 0.4 & 42.26 $\pm$ 0.04 & & & & \\
& & & 52356 & 3.3 $\pm$ 0.4 & 42.83 $\pm$ 0.04 & & & & \\
& & & 52366 & 10.8 $\pm$ 0.8 & 43.33 $\pm$ 0.03 & & & & \\
& & & 55653 & 55.2 $\pm$ 3.0 & 43.32 $\pm$ 0.02 & & & & \\
\rule{0pt}{2ex}  

& & Mg~II & 52347 & 1.7 $\pm$ 0.1 & 42.85 $\pm$ 0.03 & & & & \\
& & & 52355 & 5.3 $\pm$ 0.3 & 43.30 $\pm$ 0.02 & & & & \\
& & & 52356 & 15.8 $\pm$ 1.7 & 43.39 $\pm$ 0.05 & & & & \\
& & & 52366 & 8.6 $\pm$ 0.6 & 43.16 $\pm$ 0.03 & & & & \\
& & & 55653 & 72.9 $\pm$ 4.6 & 43.37 $\pm$ 0.03 & 2739 $\pm$ 224 & 8.18 $\pm$ 0.10 & -0.30 $\pm$ 0.20 & -1.70 $\pm$ 0.10 \\
\rule{0pt}{2ex}  

J125032.57+021632.1 & 0.959$^\textrm{b}$ & Mg~II & 52024 & 3.6 $\pm$ 0.4 & 42.29 $\pm$ 0.05 & & &  & \\
& & & 55631 & 8.2 $\pm$ 0.2 & 42.54 $\pm$ 0.01 & 4360 $\pm$ 213 & 8.06 $\pm$ 0.08 & -1.01 $\pm$ 0.19 & -2.41 $\pm$ 0.08  \\
\rule{0pt}{2ex}  

J130823.70+354637.0 & 1.055$^\textrm{c}$ & Mg~II & 55335 & 8.5 $\pm$ 1.2 & 42.46 $\pm$ 0.06 & 2250 $\pm$ 718 & 7.43 $\pm$ 0.29 & -0.46 $\pm$ 0.34 & -1.86 $\pm$ 0.30 \\
& & & 55597 & 4.9 $\pm$ 0.7 & 42.74 $\pm$ 0.06 & & & & \\
\rule{0pt}{2ex}  

J143758.67+300207.1 & 1.230$^\textrm{b}$ & Mg~II & 53757 & 0.4 $\pm$ 0.2 & 42.14 $\pm$ 0.02 & & & & \\
& & & 55277 & 25.6 $\pm$ 0.5 & 42.76 $\pm$ 0.01 & 3829 $\pm$ 281 & 8.08 $\pm$ 0.09 & -0.81 $\pm$ 0.19 & -2.21 $\pm$ 0.09 \\
\rule{0pt}{2ex}  

J220643.28-003102.5  & 1.051$^\textrm{d}$ & Mg~II & 52937 & 9.9 $\pm$ 1.5 & 42.81 $\pm$ 0.06 & 4189 $\pm$ 1683 & 8.19 $\pm$ 0.36 & -0.87 $\pm$ 0.40 & -2.27 $\pm$ 0.36 \\
& & & 55481 & 4.9 $\pm$ 0.8 & 42.61 $\pm$ 0.07 & & & & 

\enddata
\tablenotetext{}{The statistical uncertainties on the fitted parameters (line rest-frame EW, luminosity, and FWHMs) are estimated from simulating the noise in each pixel of each spectrum (see Section 2.2). The uncertainties on the derived quantities ($M_{\rm{BH}}$, $L_{\rm{bol}}$/$L_{\rm{Edd}}$, $L_{\rm{BLR}}$/$L_{\rm{Edd}}$) include the propagated statistical uncertainties from the fitting, as well as all statistical and systematic uncertainties in the scaling relations used in Section 3.1.}
\tablenotetext{a}{This blazar did not transition, but is likely to be a transition blazar (see Section 4)} 
\tablenotetext{b}{New redshift.} 
\tablenotetext{c}{Redshift from Massaro et al. (2009).} 
\tablenotetext{d}{Redshift from Shaw et al. (2013).} 
\end{deluxetable*}

\subsection{BEL Fitting and the Transition Blazar Sample}

	The details of our BEL fitting closely follow \citet[][hereafter SH11]{sh11}, to allow for comparison 
of the BEL properties of our transition blazars to the general SDSS quasar population. The primary 
difference in our fitting in comparison to SH11 is that we opt not to fit templates for Fe emission since 
the continuua in transition blazars are strongly dominated by jet emission, and thus the Fe line 
complexes are not well-detected in their spectra. For C~IV, Mg~II, and H$\alpha$, we fit a power-law 
to the continuum wavelength windows used by SH11; the flux densities in the BEL wavelength
windows (also the same as in SH11) are then normalized by the fitted continuum and integrated to 
find the EW. For C~III], which was not analyzed in SH11, we use continuum windows of 
1810-1830\AA~and 1976-1996\AA, and the BEL window is 1830-1976\AA. The choice of these 
wavelength windows for C~III] is informed by the SDSS composite quasar spectrum of \citet{va01}.

	For each BEL, we fit both a broad and narrow component, each using a Gaussian with the 
same line center fixed to their lab vacuum wavelengths. However, the higher-ionization C~IV 
line is well-known to exhibit intrinsic blueshifts \citep{ga82, ri02, ri11}, and so we allow the line center 
(assumed to be the same for both broad and narrow components) to be a single free parameter.
The FWHM of the narrow component is constrained to be $<$1200 km s$^{-1}$, and the 
amplitudes of all components are constrained to be positive. For H$\alpha$, we fit 
additional Gaussians for the narrow [N~II] $\lambda$$\lambda$6548, 6584 lines, as well as 
[S~II] $\lambda$$\lambda$6717, 6731. We visually inspect the fit of each emission line and 
continuum region to ensure that it was not strongly affected by problematic pixels or absorption 
features. The statistical uncertainties on all quantities derived from the fitting (notably the EWs, 
luminosities, and FWHMs) are estimated by simulating the noise in each spectrum, by resampling 
each pixel from a Gaussian distribution with $\sigma$ set to the uncertainty in the pixel flux density. 
All pixels in each spectrum are resampled in this fashion and refitted with the above procedure 
10$^3$ times; the uncertainties on the fitted quantities are then based on their 1$\sigma$ range 
in the simulated spectra. Several studies have indicated a likely dependence of
quasar BEL FWHMs on orientation \citep{wi86, ja06, fi11, ru13}, which may bias our FWHM 
measurements and black hole mass estimates due to the strong jet-alignment of blazars.
However, this bias cannot be significantly larger than the $\lesssim$0.2 dex dispersion in the
log-normal distribution of BEL FWHMs of quasars \citep{sh08}, and thus
cannot account for the 1-2 dex difference in the Eddington ratio between our transition 
blazars (see Section 3.1) and that typical of radiatively inefficient accretion.

	We identify six blazars for which any BEL in the spectrum crossed the 5\AA~EW threshold in the 
rest-frame, transitioning between the canonical BLL and FSRQ classifications (or vice versa). 
The properties of these transition blazars are listed in Table 1 (and discussed in Section 3), and 
Figures 2-8 show their multi-epoch spectra, the fitting of the continuum-normalized BEL region in each 
of the epochs, as well as their long-term photometric light curves (discussed in Section 3.4). The observed changes 
in the EW in the six blazars across the 5\AA~threshold vary wildly, up to a factor of $>$60 for 
SDSS J101603.13+051302.3. Due to the small sample size, we will not attempt to investigate the 
incidence of these blazar subclass transitions in our current study, and instead focus on understanding 
the nature of this phenomenon.

	Since transition blazars intermittently show small BEL EWs by 
definition, redshifts can be difficult to obtain from spectra. Indeed, for two of these transition blazars 
as well as one additional blazar (SDSS J083223.22+491321.0) which did not transition, no previous 
redshifts have been reported. We report new secure redshifts from the SDSS pipeline using the spectral 
epochs where the BELs have the largest EWs (are are thus most well-detected). Although 
SDSS J083223.22+491321.0 did not transition, we nevertheless show its spectra and BEL fits in Figure 8 
due to its newly-determined redshift and its special implications for transition blazars (discussed in 
Section 5). We discuss the properties of each of these individual blazars in more detail in Section 4.

\begin{figure}
\begin{center}
\includegraphics[width=0.49\textwidth, height=6cm]{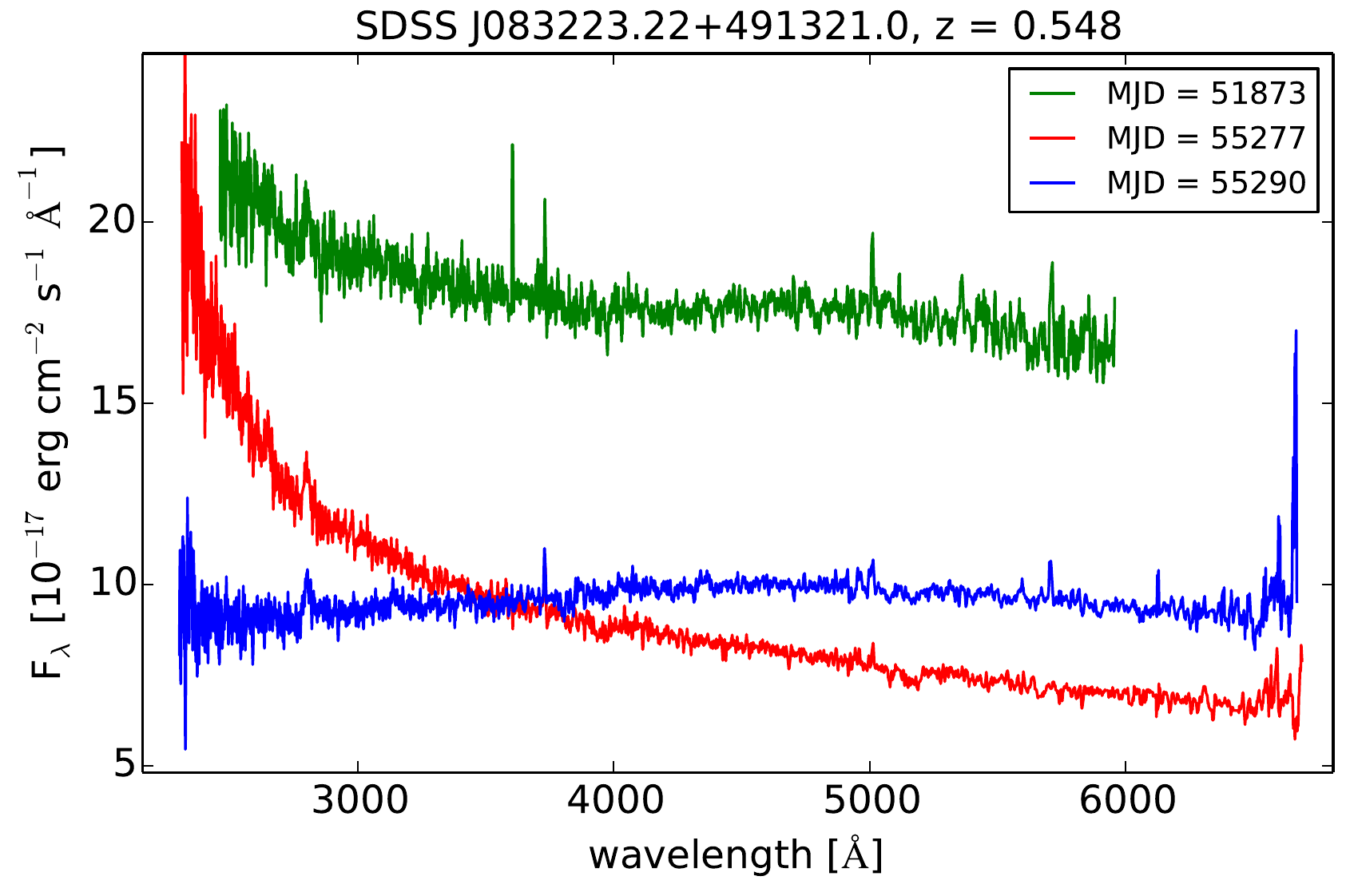} \\
\includegraphics[width=0.49\textwidth]{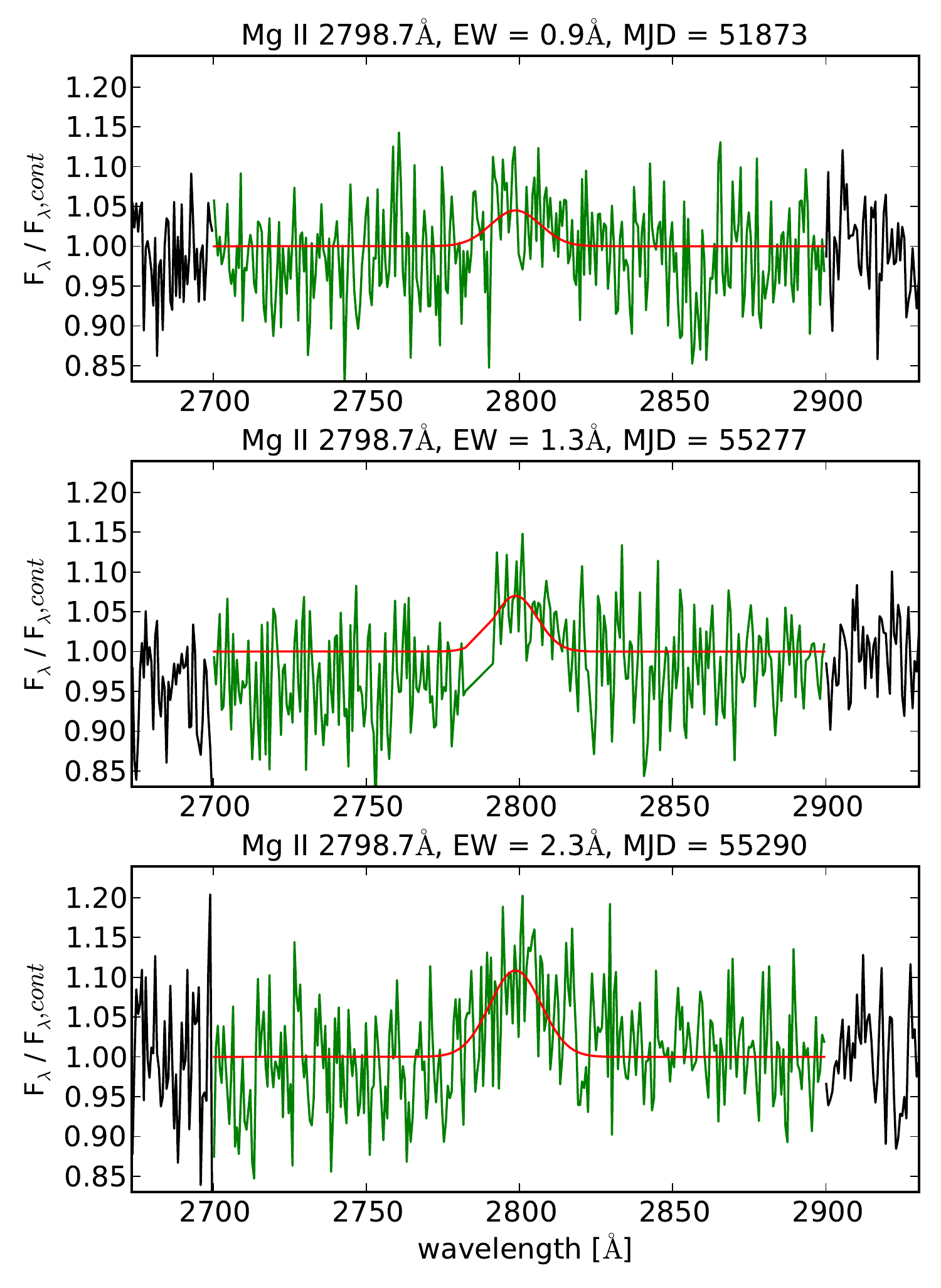} \\
\includegraphics[width=0.49\textwidth, height=5cm]{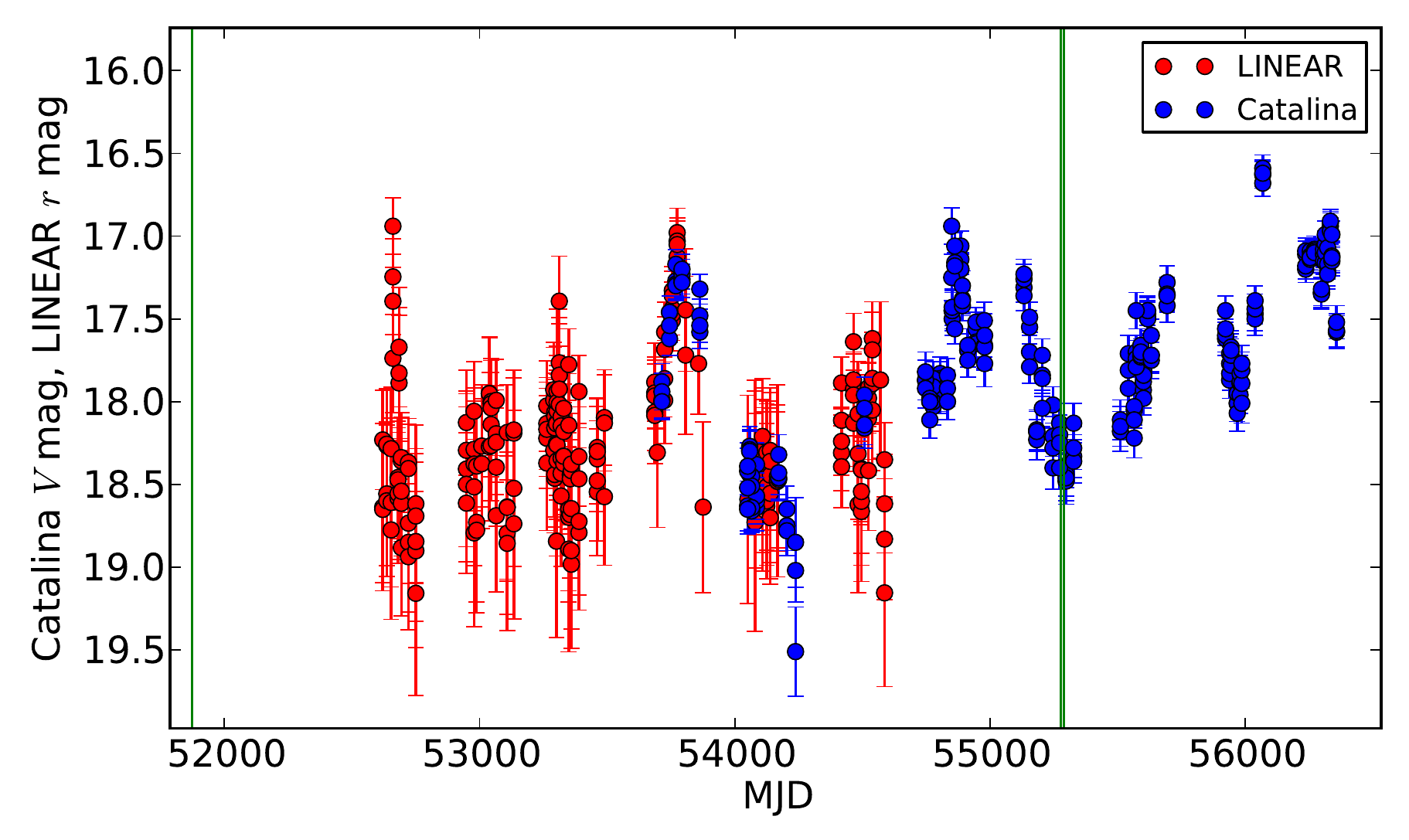} 
\caption{SDSS J083223.22+491321.0: multi-epoch observed-frame SDSS spectra, rest-frame broad emission 
line fitting, and photometric light curve. The dates of the spectroscopic epochs are indicated by the green lines 
in the light curve.}
\end{center}
\end{figure}
\begin{figure}
\begin{center}
\includegraphics[width=0.49\textwidth, height=6cm]{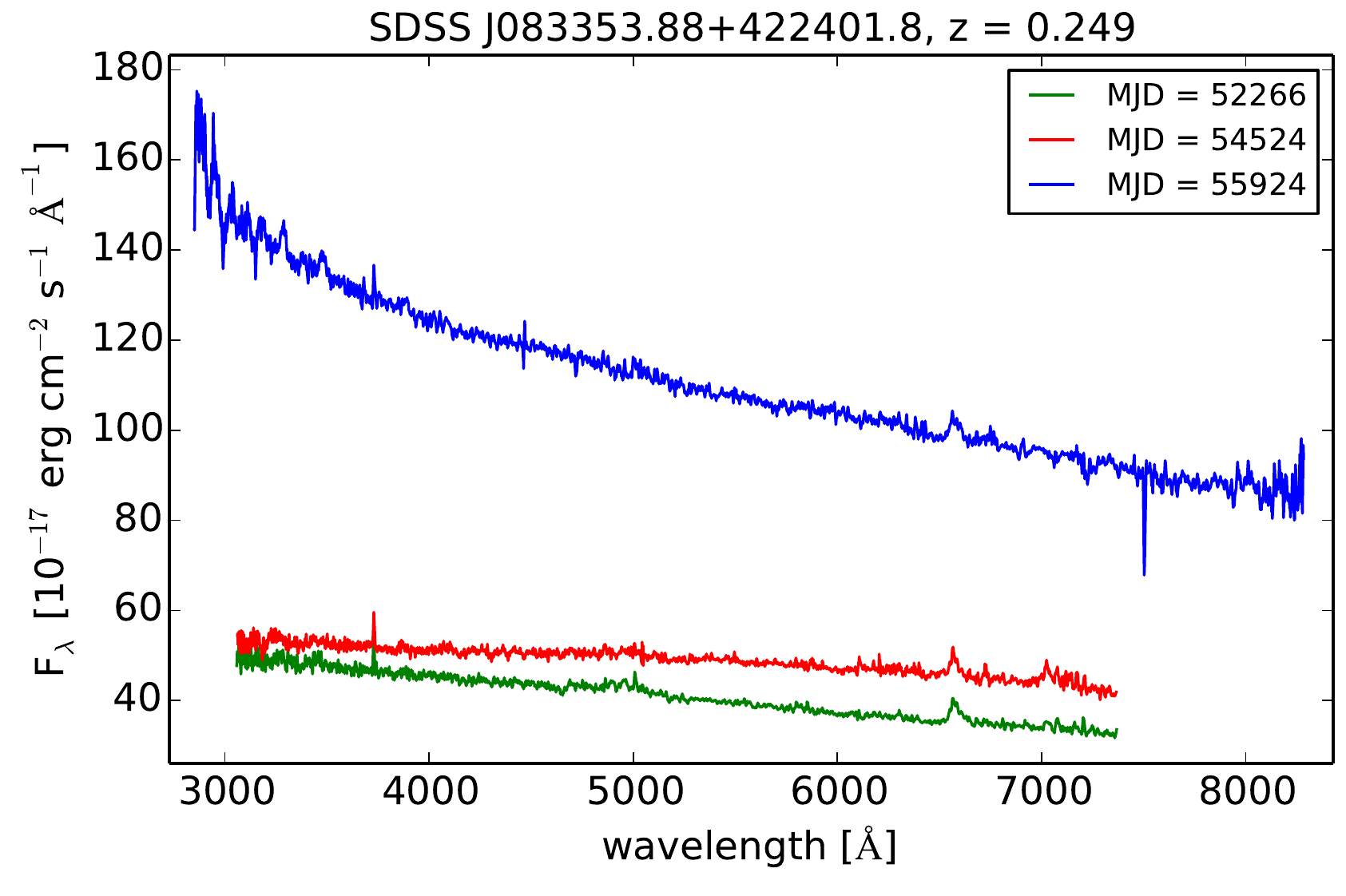} \\
\includegraphics[width=0.49\textwidth]{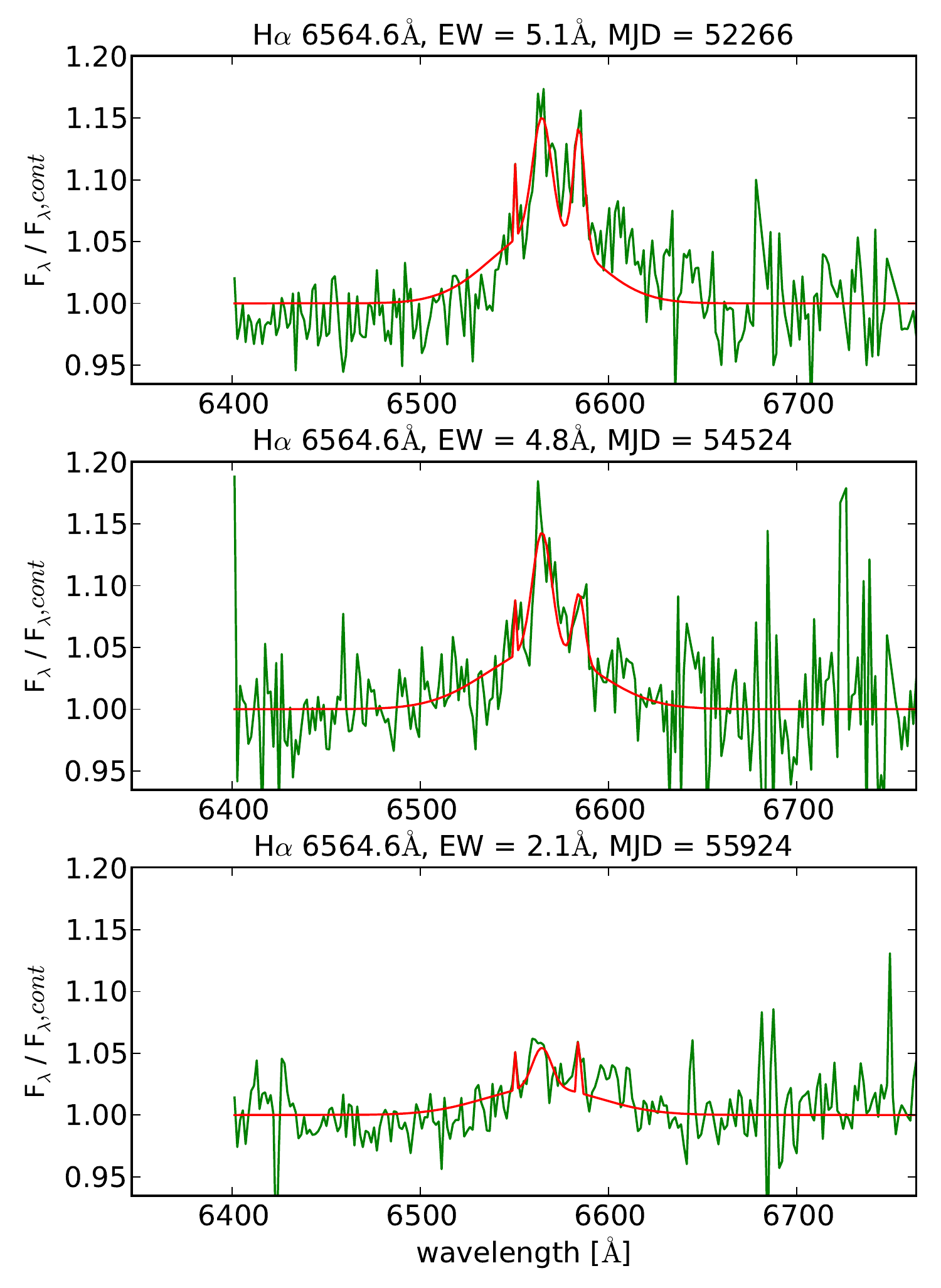} \\
\includegraphics[width=0.49\textwidth, height=5cm]{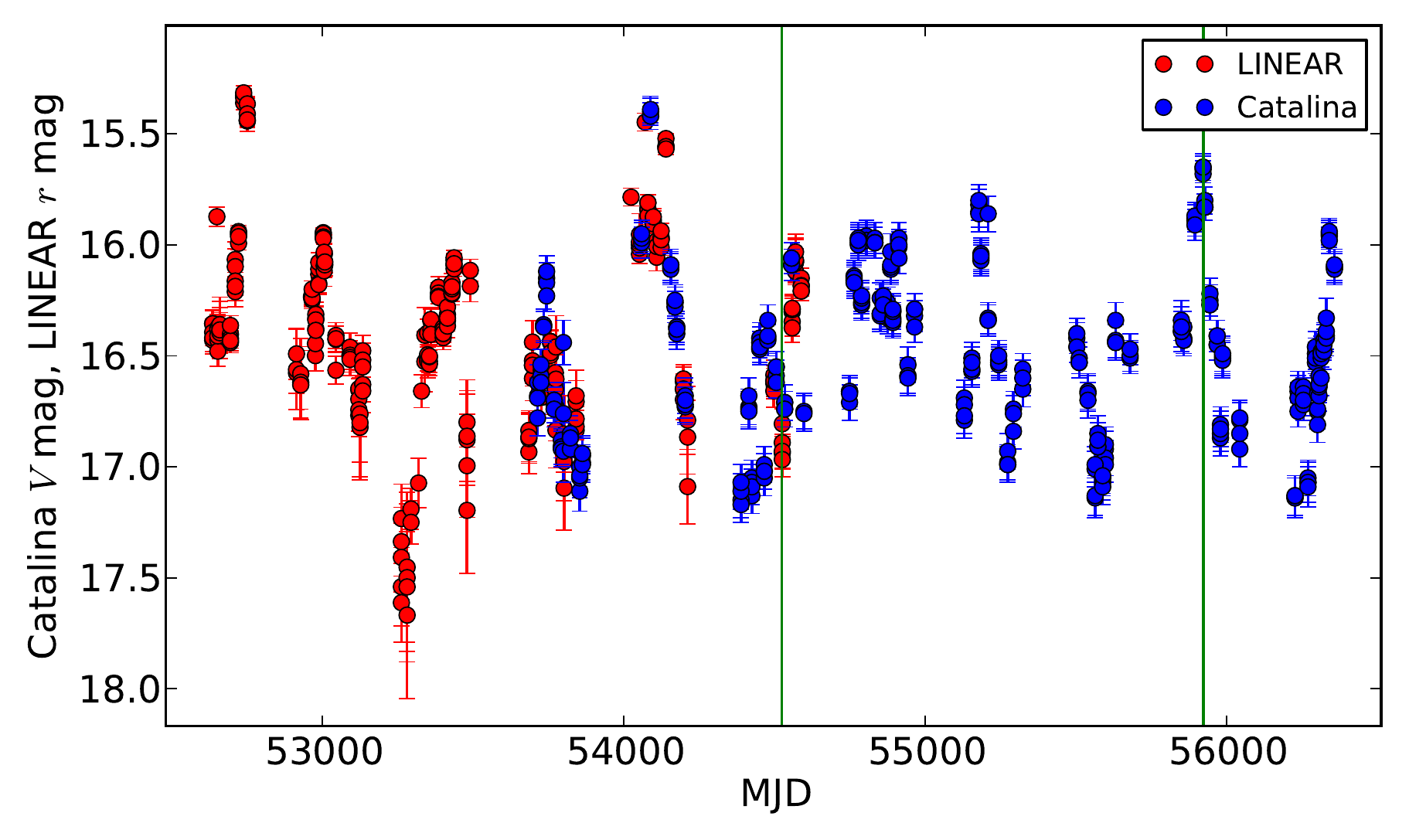} 
\caption{SDSS J083353.88+42241.8, similar to Figure 2.}
\end{center}
\end{figure}
\begin{figure*}
\begin{center}
\includegraphics[width=0.49\textwidth, height=5.5cm]{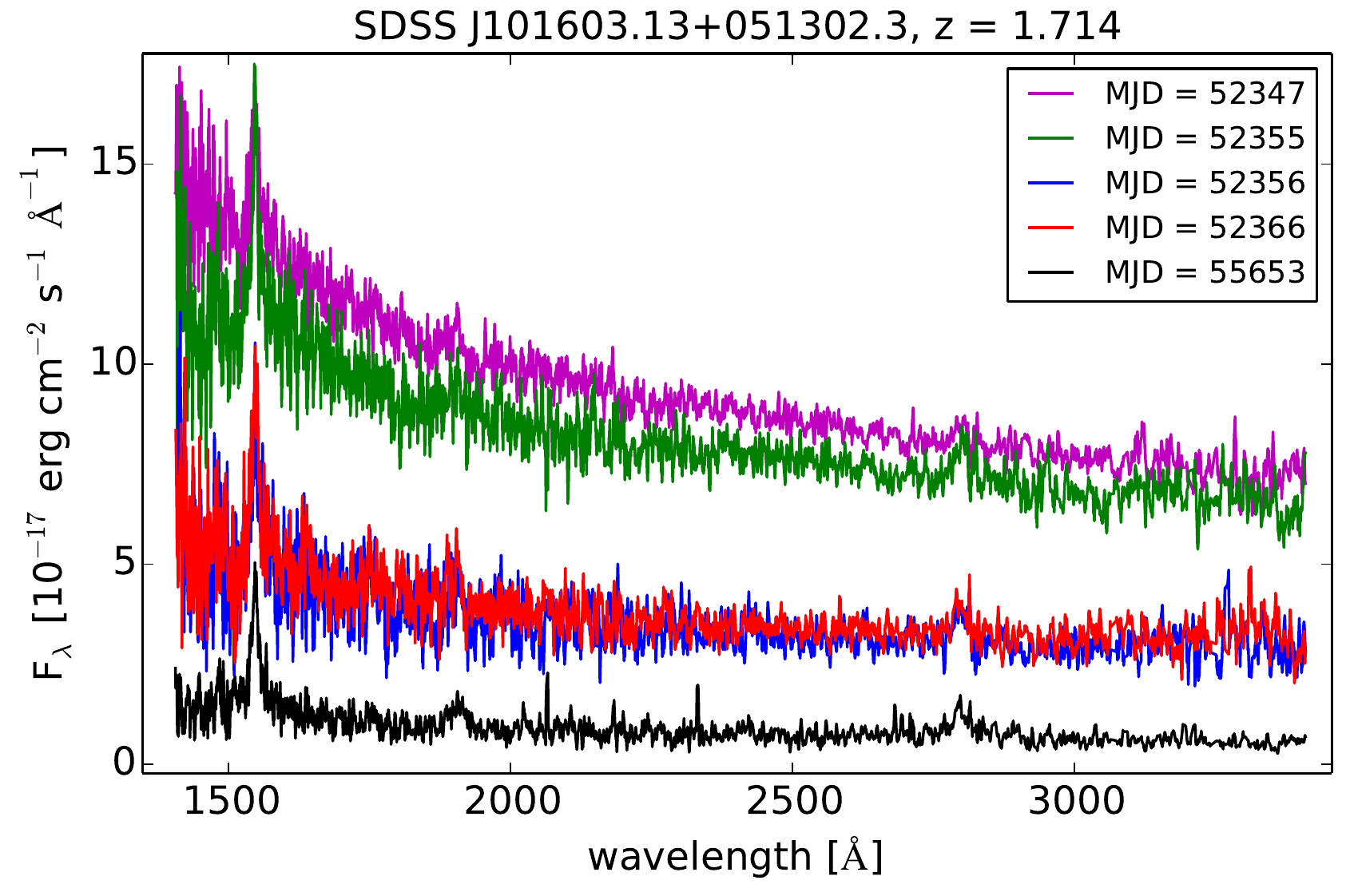} 
\includegraphics[width=0.49\textwidth, height=5.5cm]{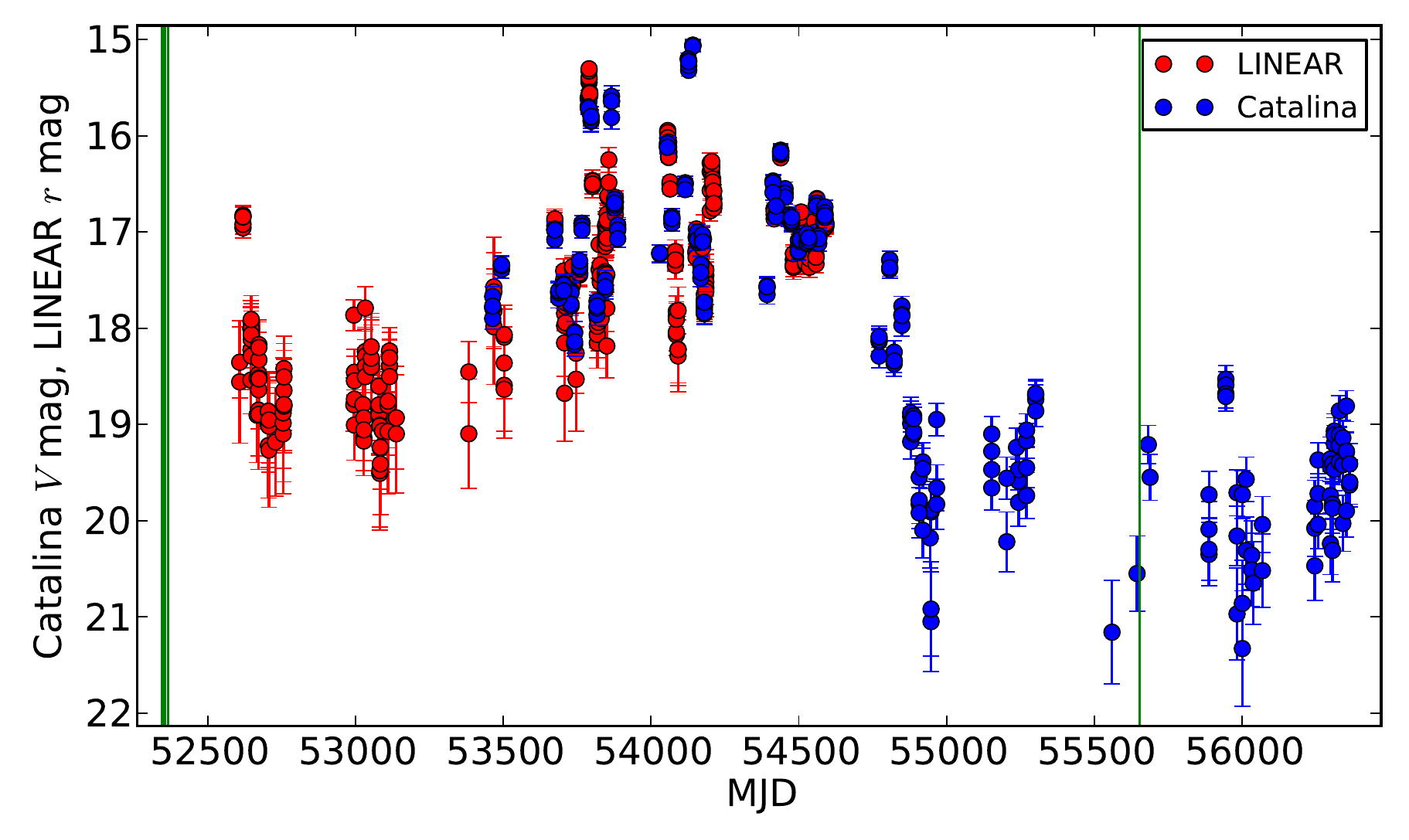} \\
\includegraphics[width=.99\textwidth]{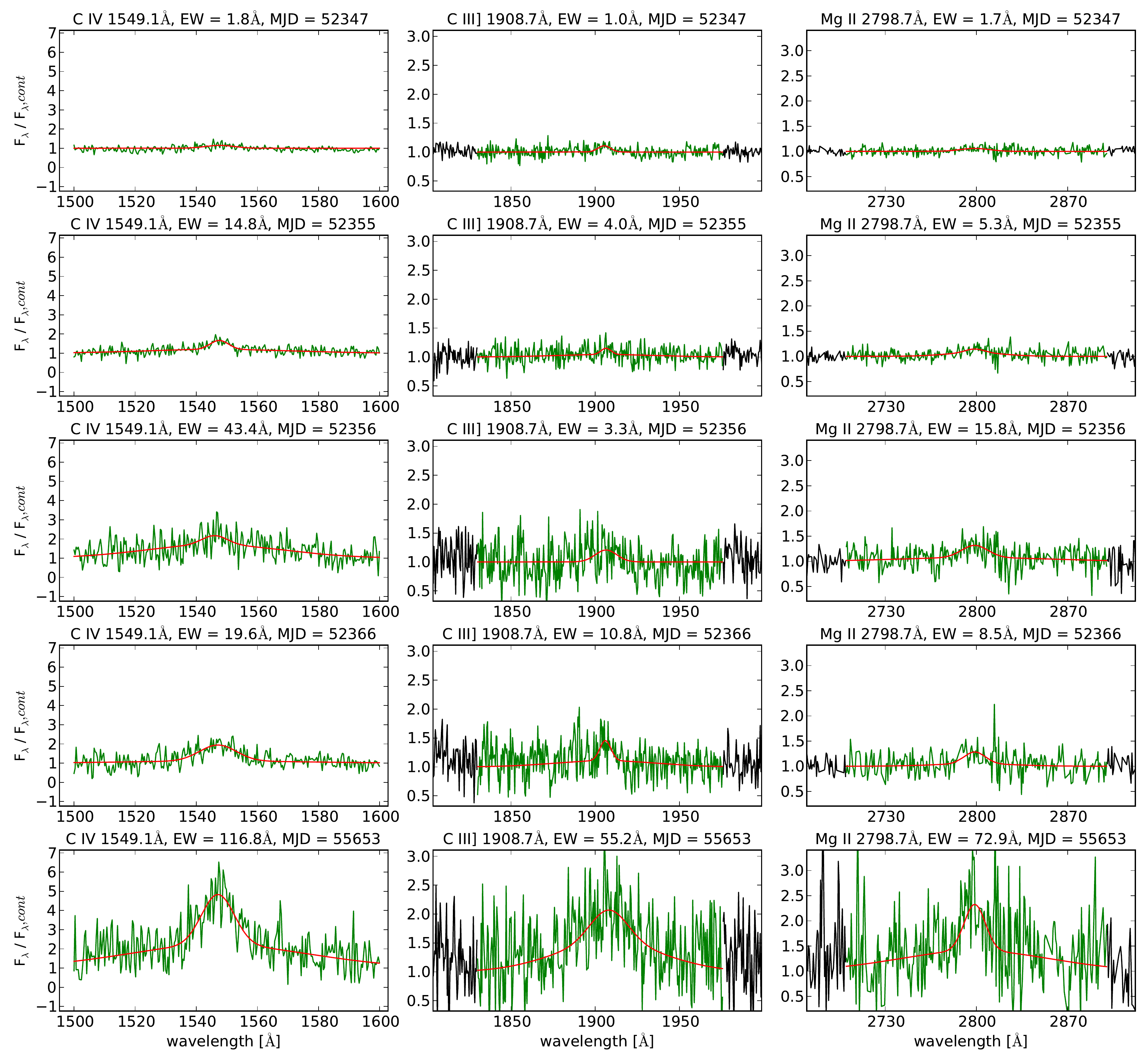} 
\caption{SDSS J10163.13+05132.3, similar to Figure 2.}
\end{center}
\end{figure*}
\begin{figure}
\begin{center}
\includegraphics[width=0.49\textwidth, height=6cm]{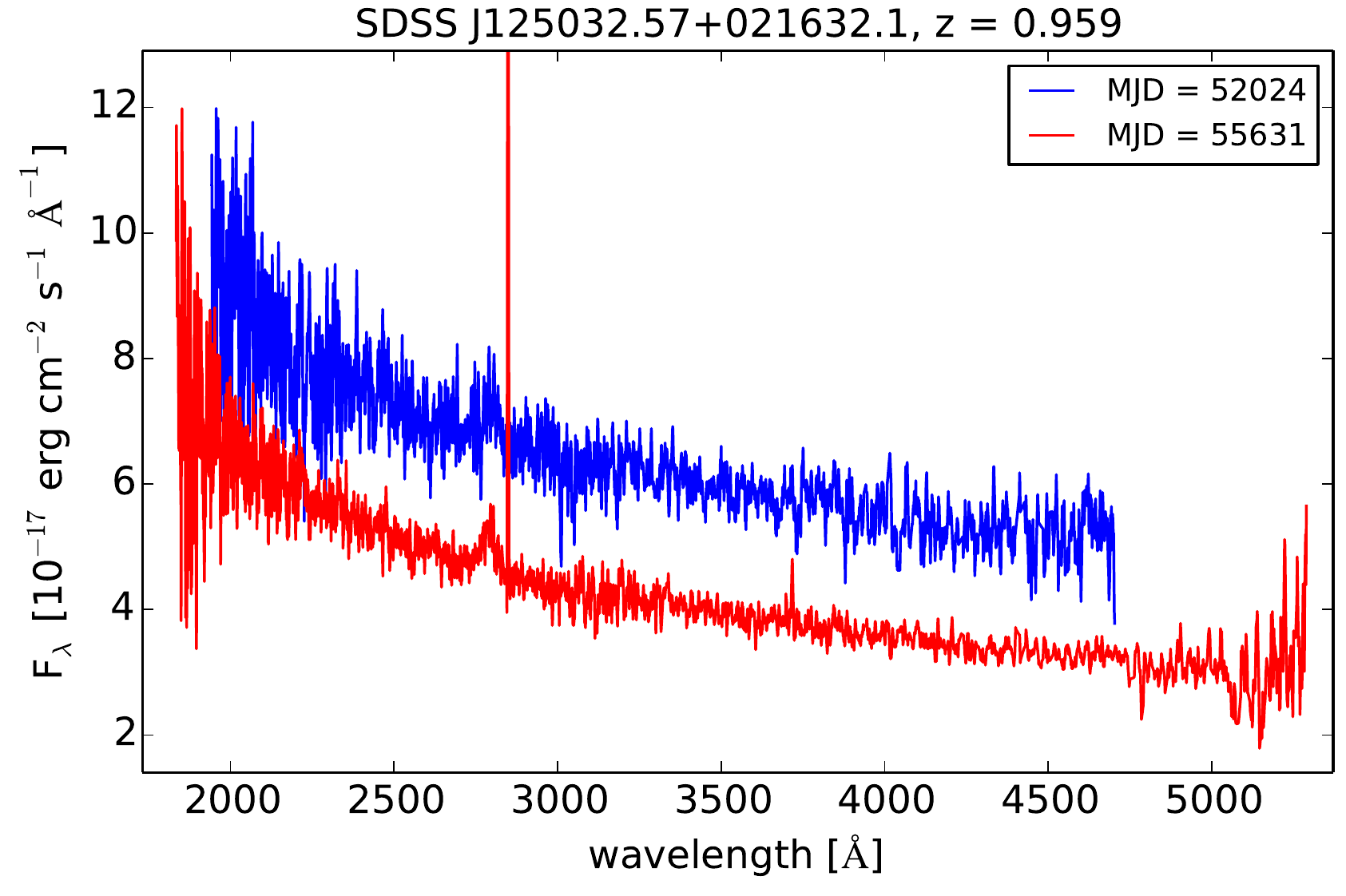} \\
\includegraphics[width=0.49\textwidth]{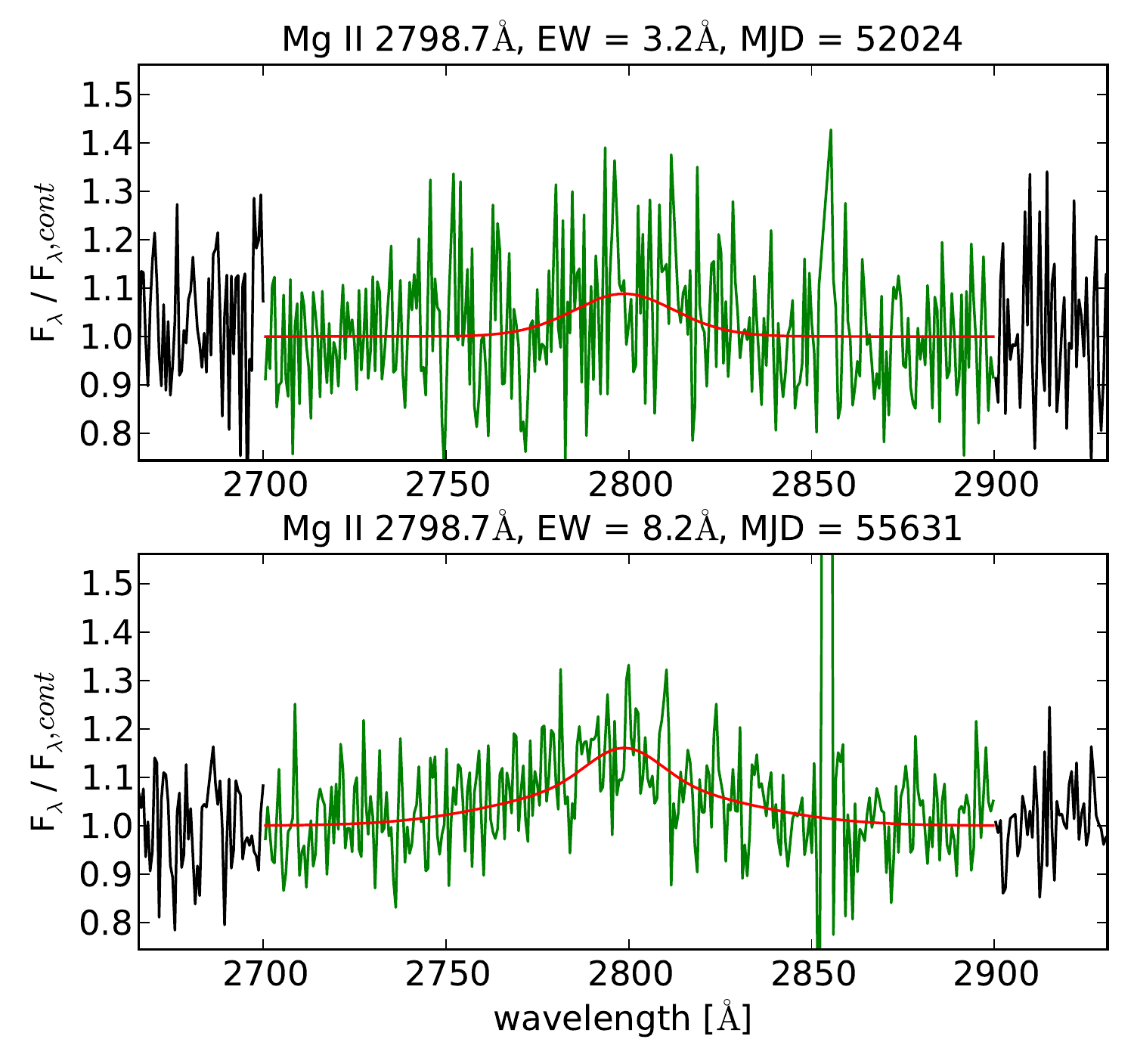} \\
\includegraphics[width=0.49\textwidth, height=5cm]{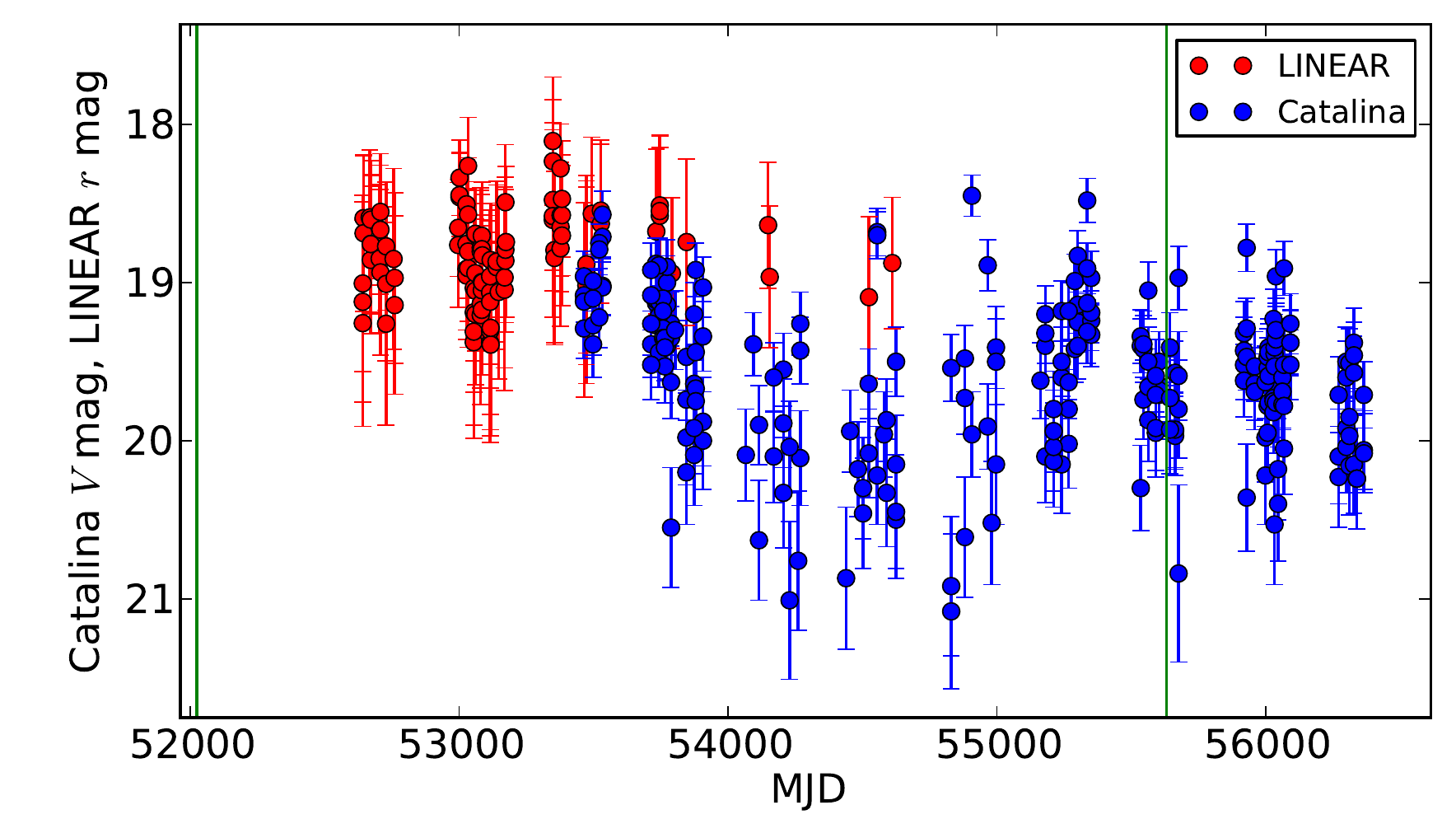} 
\caption{SDSS J125032.57+021632.1, similar to Figure 2.}
\end{center}
\end{figure}
\begin{figure}
\begin{center}
\includegraphics[width=0.49\textwidth, height=6cm]{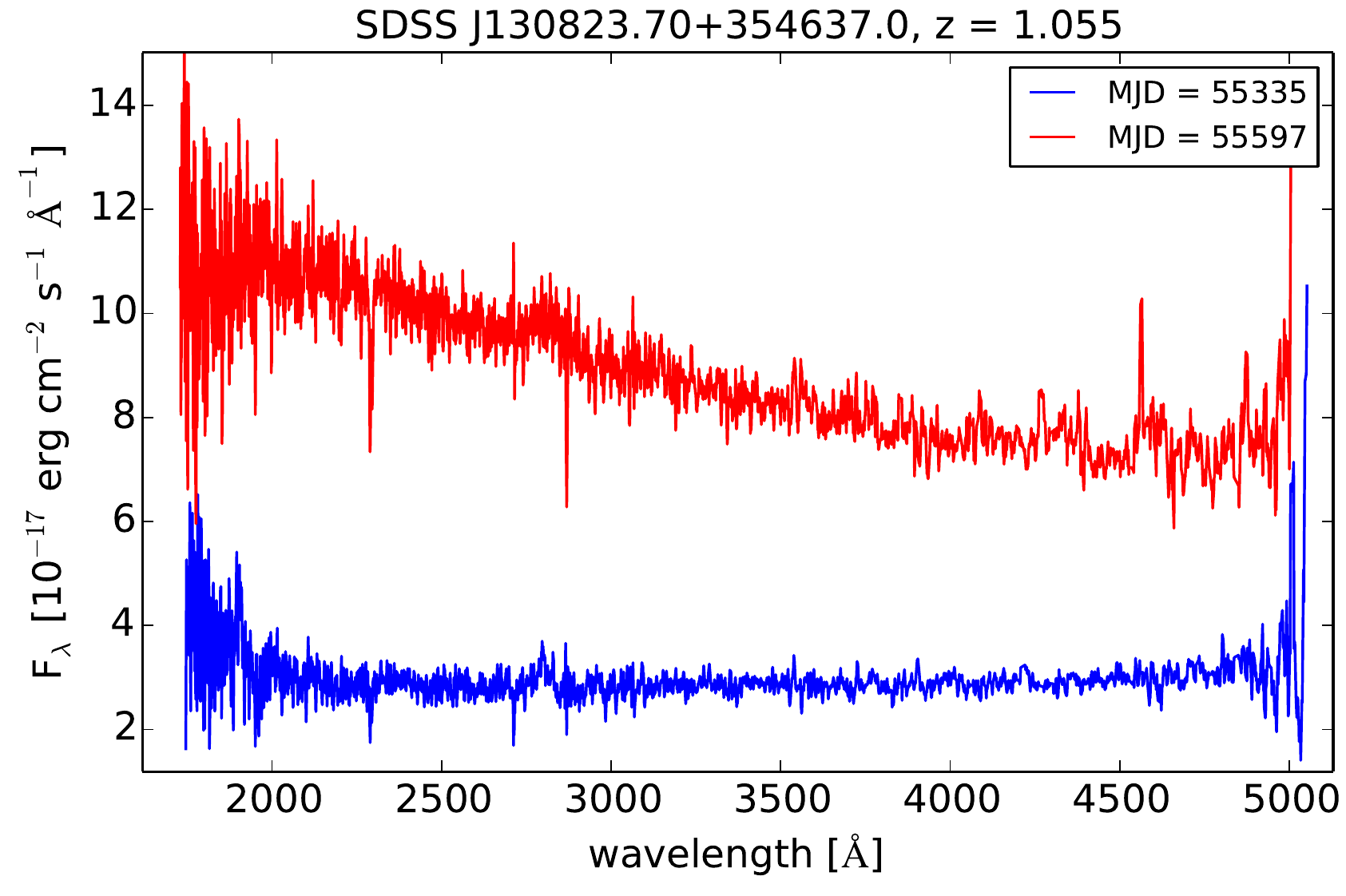} \\
\includegraphics[width=0.49\textwidth]{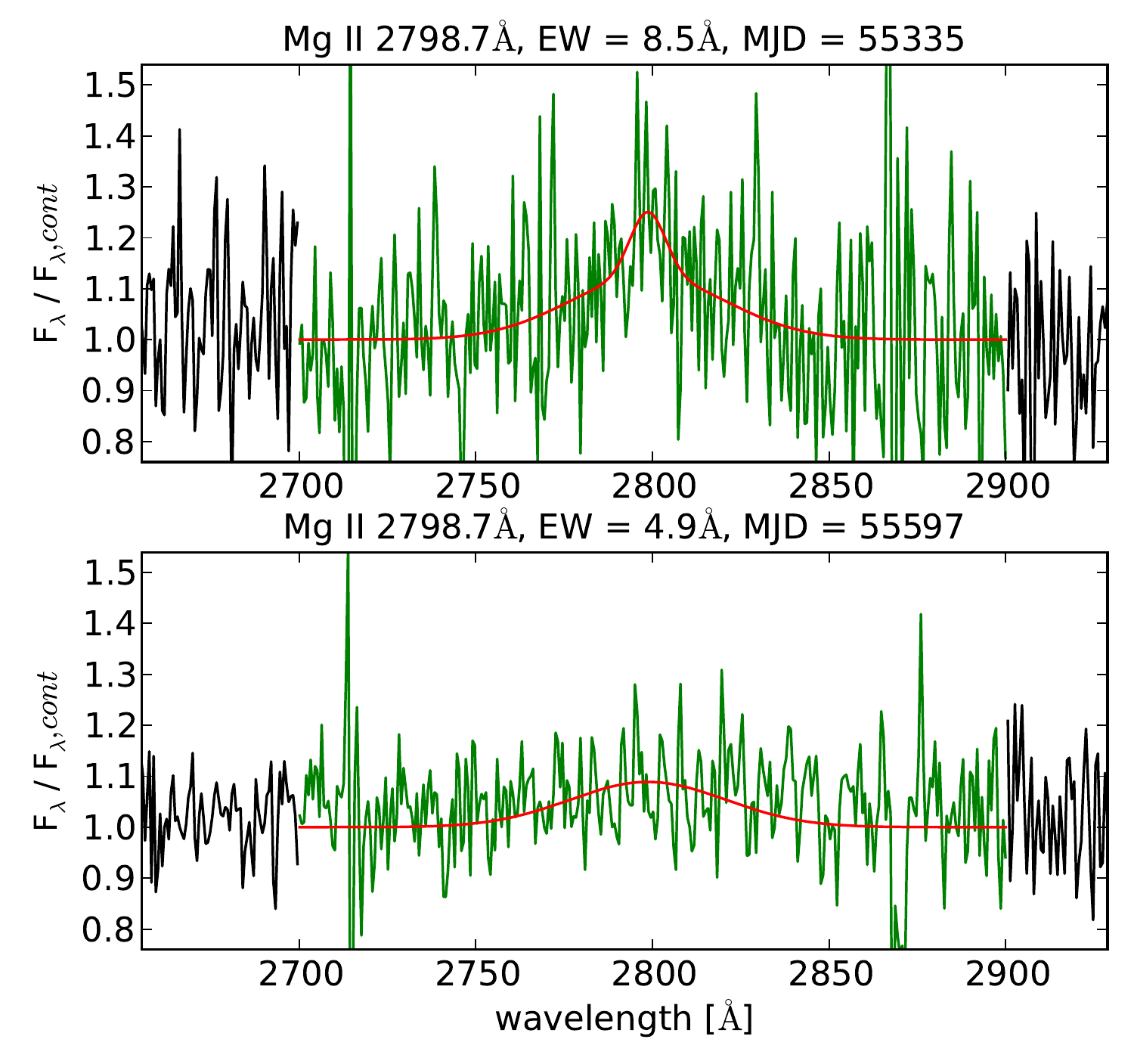} \\
\includegraphics[width=0.49\textwidth, height=5cm]{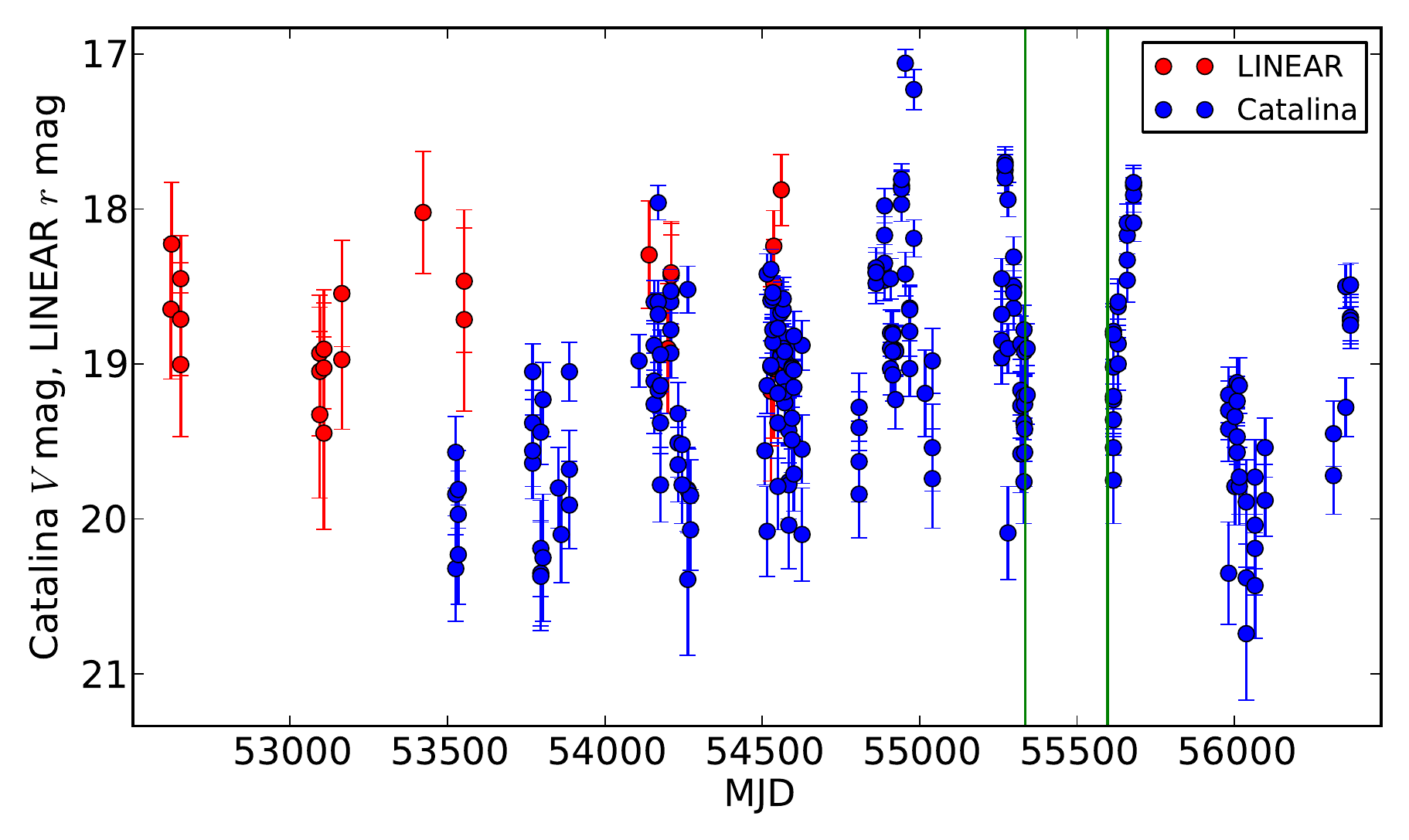} 
\caption{SDSS J130823.70+354637.0, similar to Figure 2.}
\end{center}
\end{figure}
\begin{figure}
\begin{center}
\includegraphics[width=0.49\textwidth, height=6cm]{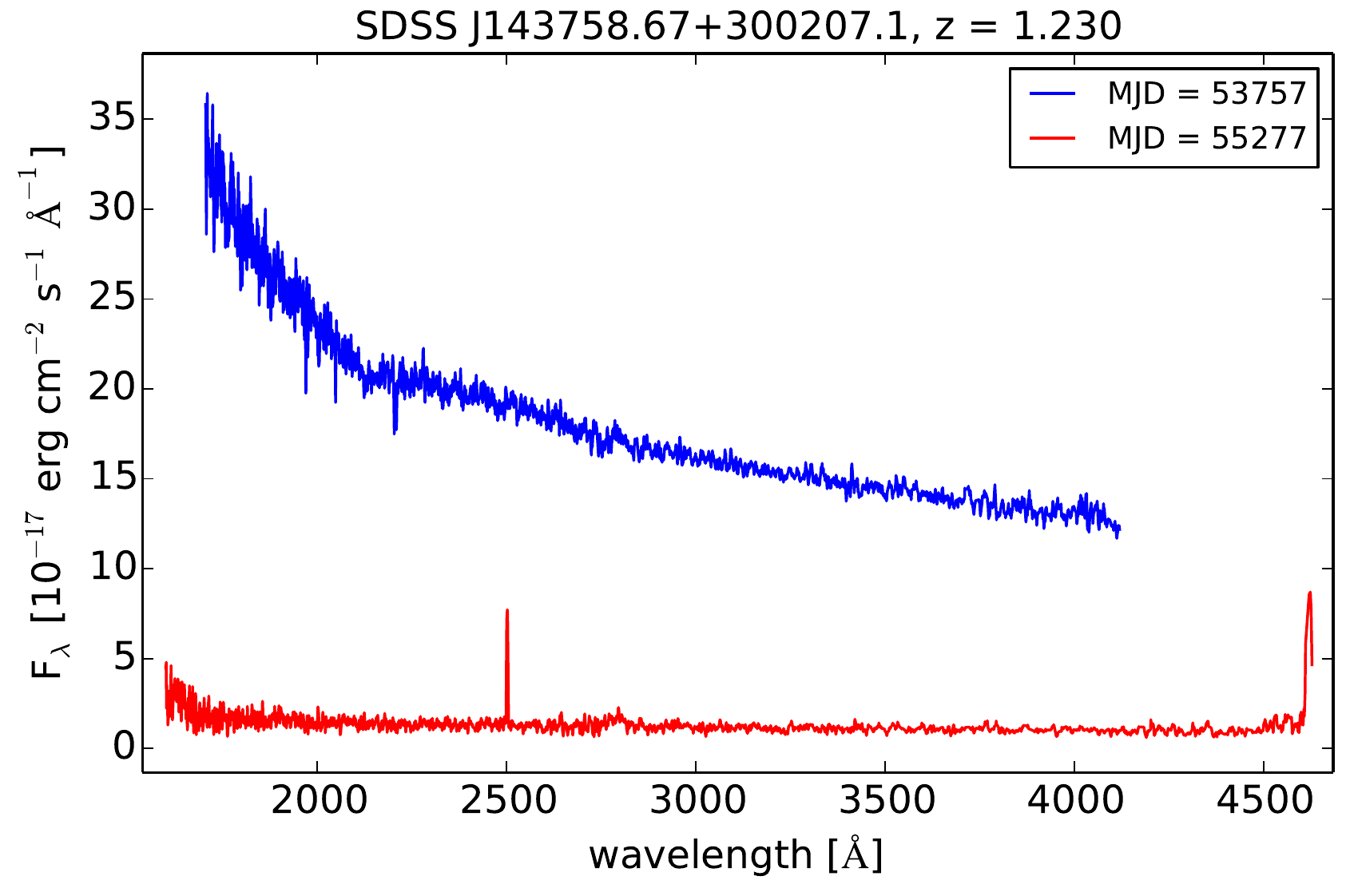} \\
\includegraphics[width=0.49\textwidth]{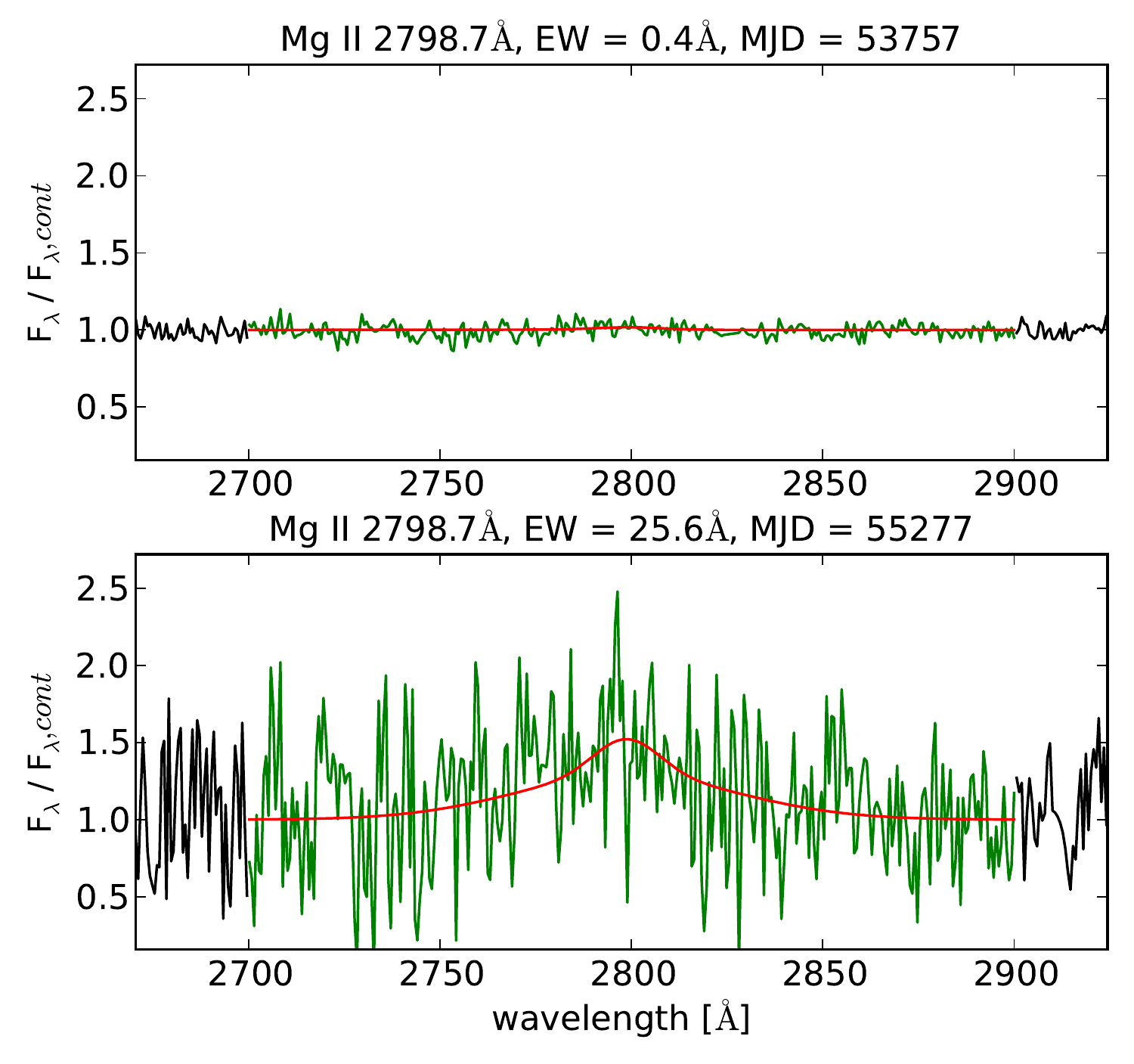} \\
\includegraphics[width=0.49\textwidth, height=5cm]{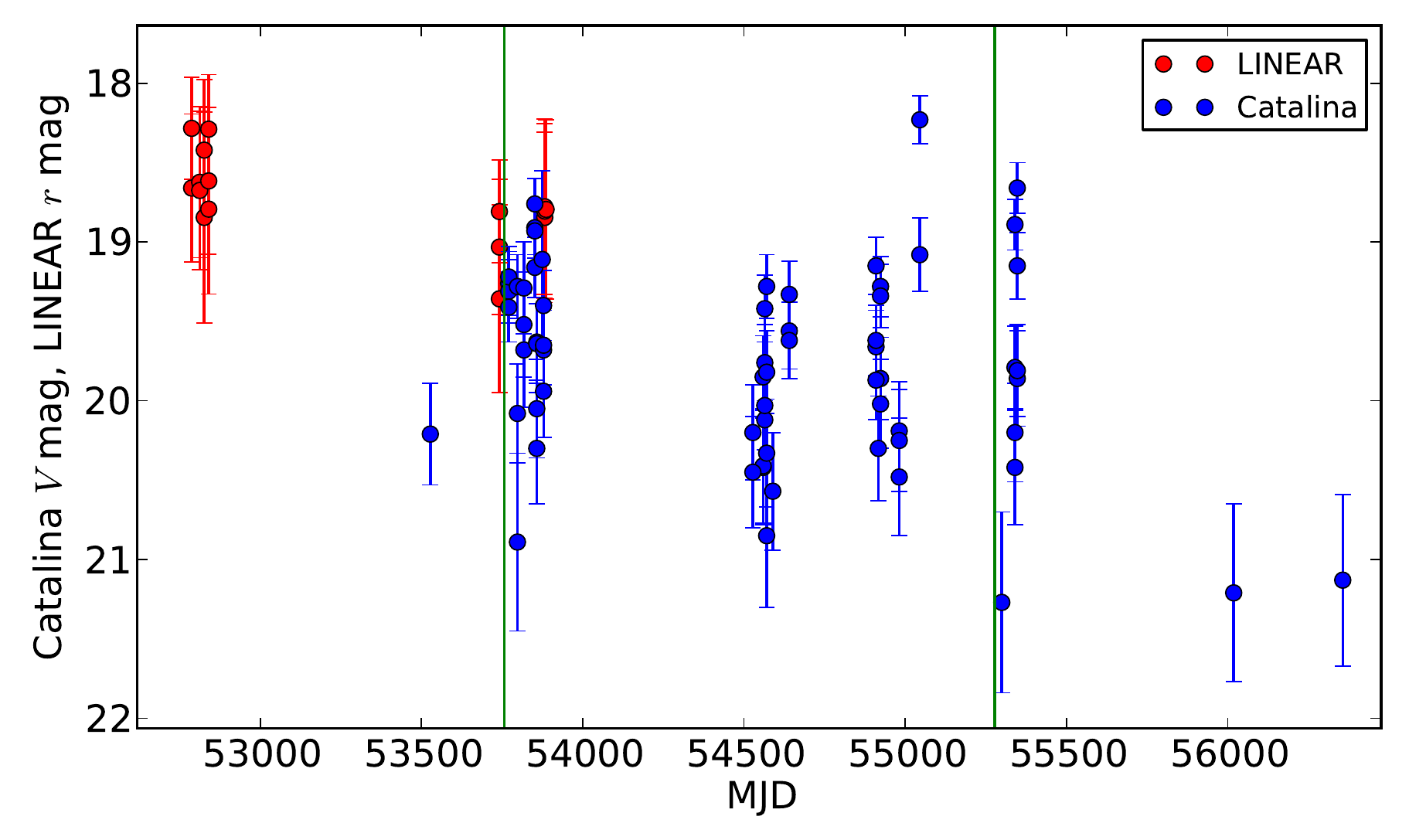} 
\caption{SDSS J143758.67+30027.1, similar to Figure 2.}
\end{center}
\end{figure}
\begin{figure}
\begin{center}
\includegraphics[width=0.49\textwidth, height=6cm]{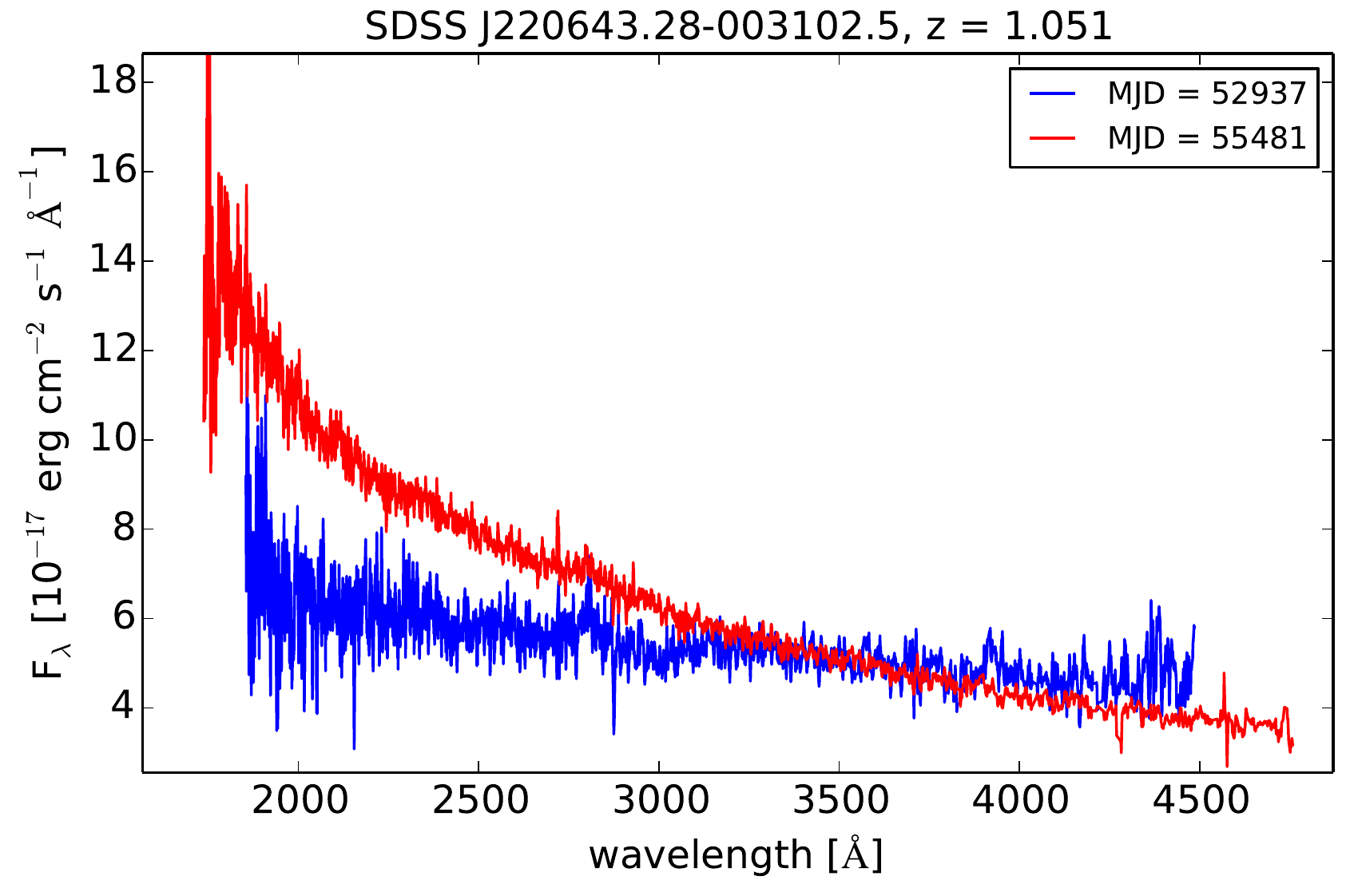} \\
\includegraphics[width=0.49\textwidth]{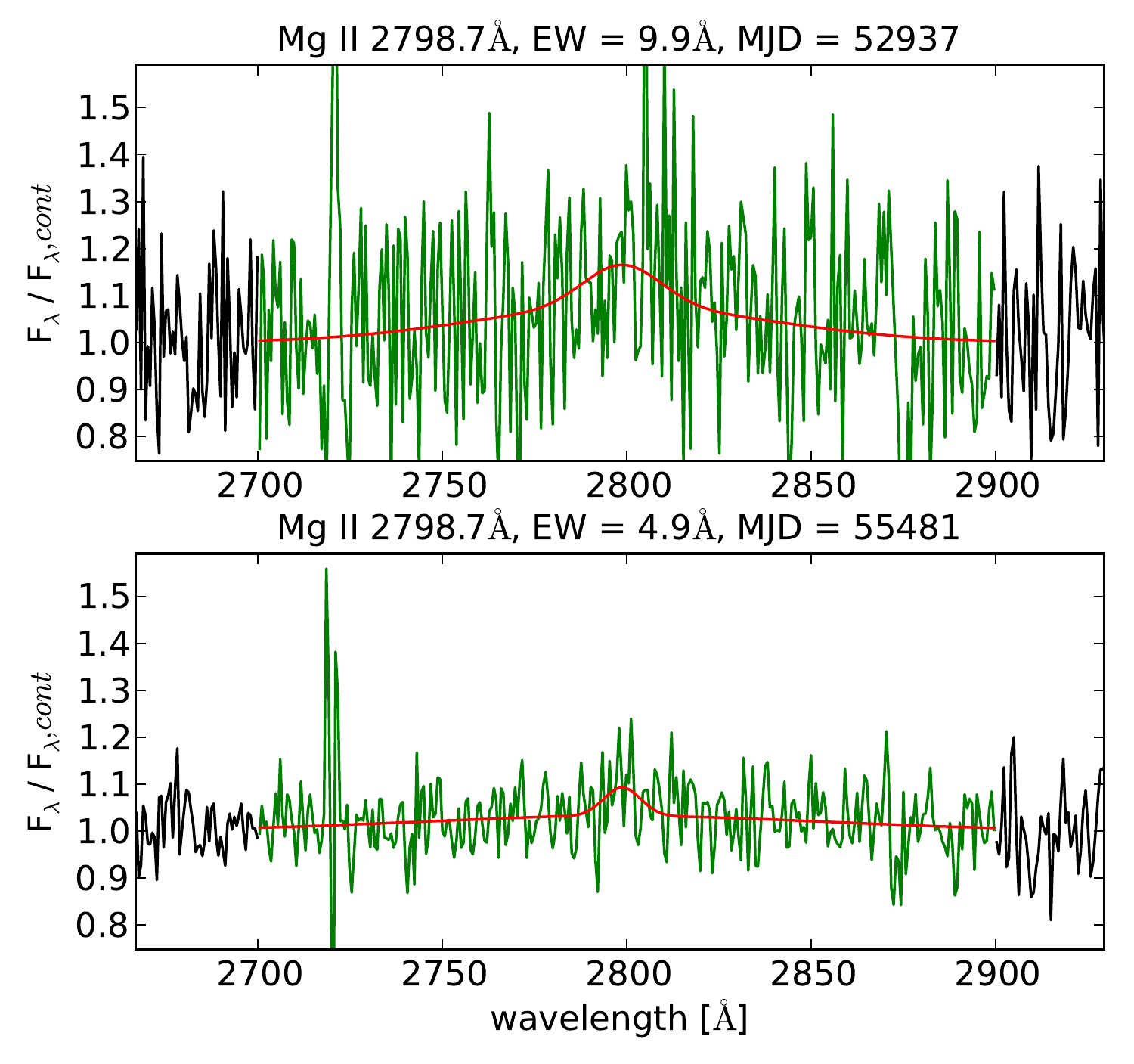} \\
\includegraphics[width=0.49\textwidth, height=5cm]{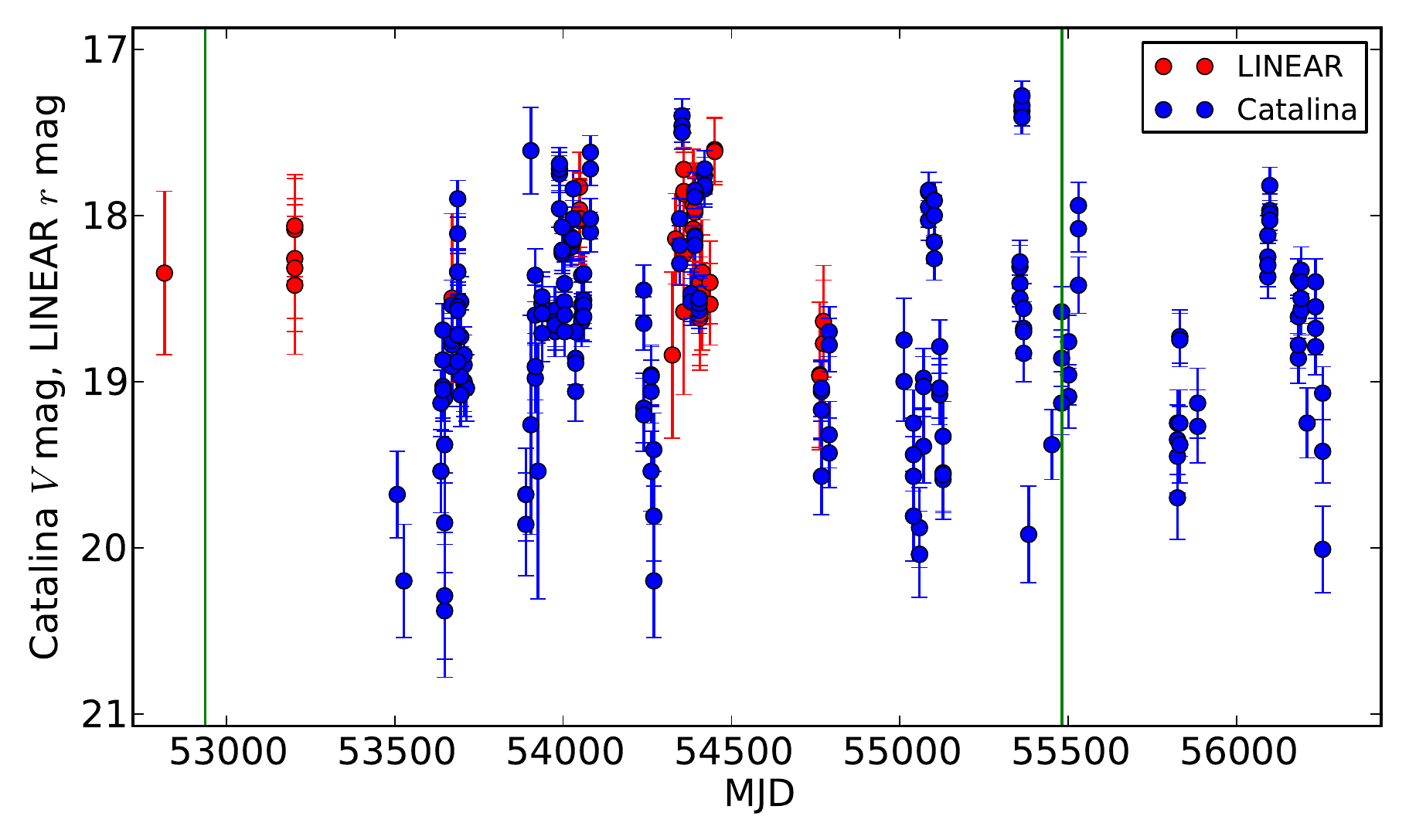} 
\caption{SDSS J220643.28-00312.5, similar to Figure 2.}
\end{center}
\end{figure}
\begin{figure*}[!t]
\begin{center}
\includegraphics[width=0.49\textwidth]{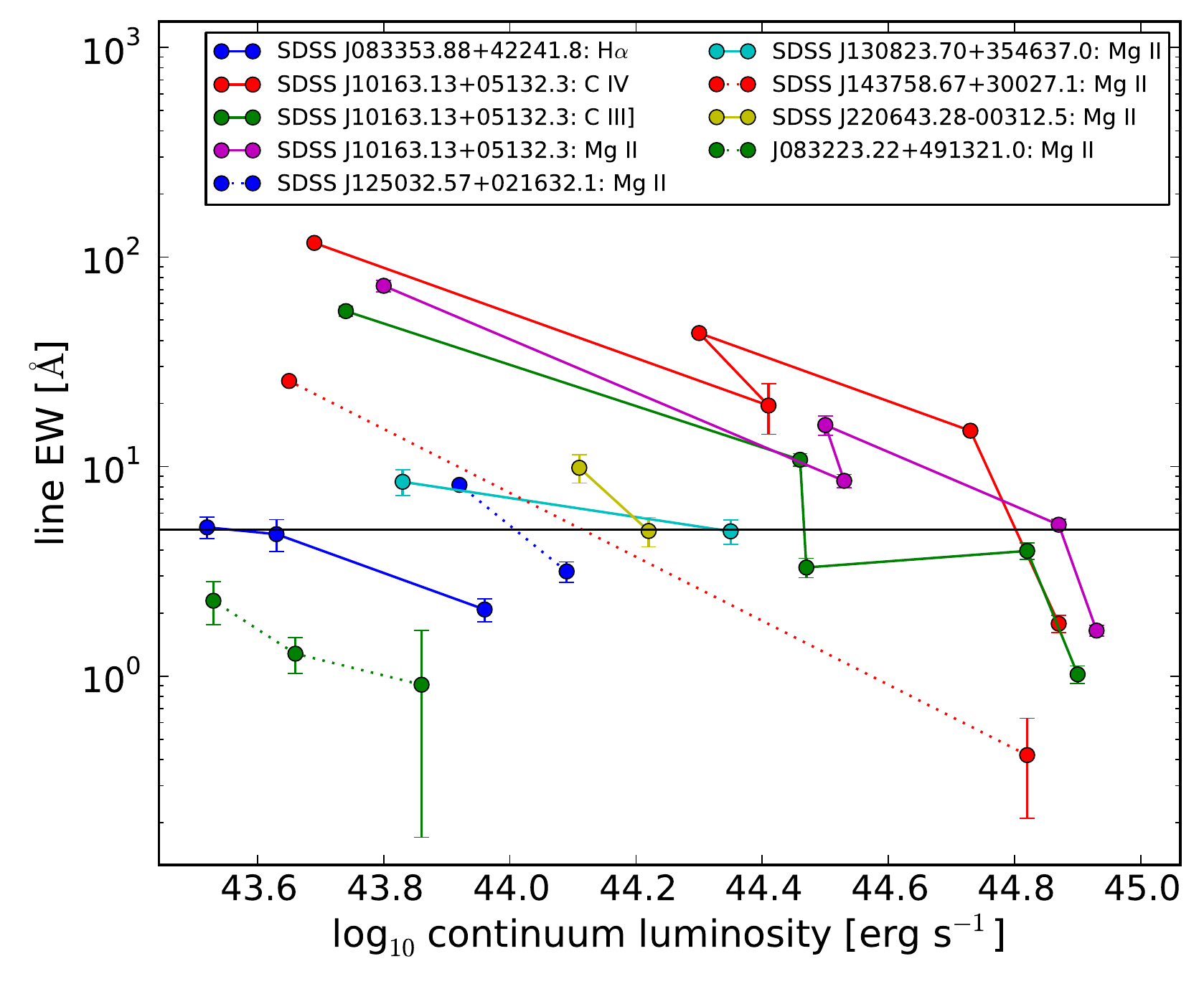}
\includegraphics[width=0.49\textwidth]{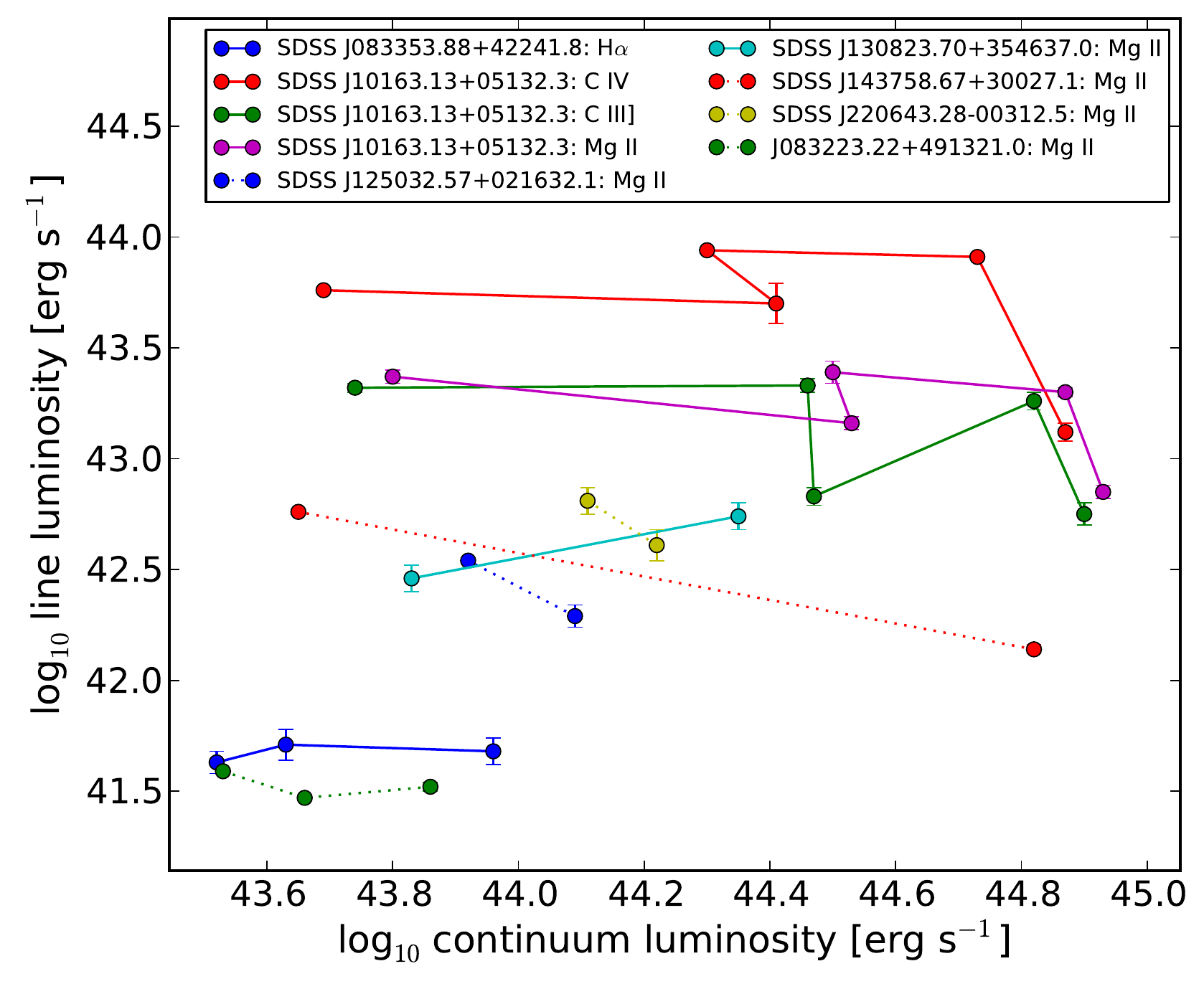} 
\caption{Left: The broad emission line rest-frame equivalent width versus the fitted continuum luminosity at the line, 
for all lines present in the six transition blazars and one additional blazar in our sample. Measurements from all 
spectral epochs for each line are shown, and connected in temporal order. The distinct anti-correlation 
strongly indicates that changes in the equivalent width of the lines are primarily due to swamping from the highly variable 
continuum. Right: Similar to the left panel, but for the line luminosity versus the continuum luminosity at the line.
The line luminosities also show variability, possibly due to variability in the ionizing continuum from a radiatively-
efficient accretion disk.}
\end{center}
\end{figure*}

\section{Properties of Transition Blazars}
\subsection{Emission Line Properties}

	The homogeneous nature of SDSS spectra allows us not only to identify a sample of transition
blazars with well-determined redshifts, but also to characterize robustly their emission line properties.
Using the BEL fits described in Section 2.1, we calculate their intrinsic line luminosities over their 
corresponding BEL wavelength windows; these line luminosities are also listed in Table 1. Figure 9 
(left panel) shows the BEL EW versus the fitted continuum luminosities in the BEL wavelength windows, 
for all epochs of the BELs of each transition blazar in temporal order. A clear anti-correlation is present 
across different BELs and in each of the different blazars, for nearly all epochs. This anti-correlation 
strongly suggests that the EW variability causing the transition blazar phenomenon in our sample 
is primarily due to swamping of the emission lines by the jet continuum emission, which is highly 
variable due to Doppler boosting. However, the emission lines themselves can also have intrinsic 
variability in their luminosities, which could contribute to their observed EW changes. Figure 9 
(right panel) shows the line luminosity versus the continuum luminosity of the same lines as the left 
panel. The lines are variable up to $\gtrsim$0.5 dex in luminosity; in some instances 
(e.g. SDSS J125032.57+021632.1), the BEL luminosities appear to show an additional anti-correlation with 
the continuum luminosity (i.e. the lines become more luminous as the continuum luminosity decreases), 
further increasing the EW variability. Variations in luminosity of BELs of 0.6 dex have also been previously 
observed in the namesake BL Lac object \citep{co96}, which is a known transition blazar.

	We use the line luminosities to estimate the bolometric Eddington ratio ($L_{\rm{bol}}$/$L_{\rm{Edd}}$), 
as well as the Eddington ratio of the broad line emission ($L_{\rm{BLR}}$/$L_{\rm{Edd}}$). These 
two quantities which probe the accretion state are often used to investigate blazar subclasses in large 
samples \citep[e.g.,][]{gh11,sh12}, allowing us to place the accretion state of our transition blazars in a 
broader context. We calculate $L_{\rm{bol}}$/$L_{\rm{Edd}}$ of the transition blazars in our sample 
following the method of \citet{sh12}, largely based on the results of SH11. We first estimate the 
unbeamed thermal continuum luminosity at 3000\AA~using its correlation with the Mg~II line 
luminosity determined by SH11 from a large sample of quasars,

\begin{equation}
	\rm{log}\left(\frac{\it{L}_{\rm{cont}}}{\rm{erg~s^{-1}}}\right) = \it{a}\cdot\rm{log}\left(\frac{\it{L}_{\rm{line}}}{erg~s^{-1}}\right) + \it{b},
\end{equation}
where $a = 1.016 \pm 0.003$ and $b = 1.22 \pm 0.11$ for the Mg~II BEL. We use this correlation to 
estimate the continuum luminosity rather than a direct measurement because the continuum emission 
in blazars is often dominated by non-thermal jet emission. The bolometric luminosity is estimated 
by multiplying the continuum luminosity at 3000\AA~from Equation 1 by the bolometric correction of $L_{\rm{bol}}$/$L_{\rm{3000}}$ = 5.15 from \citet{ri06}. Using the luminosity and the fitted FWHMs of the Mg~II lines, we estimate
black hole masses $M_{\rm{BH}}$ using the relation 

\begin{equation}
	\rm{log}\left(\frac{\it{M}_{\rm{BH}}}{\it{M}_{\odot}}\right) = \it{c} + \it{d}\cdot\rm{log}\left(\frac{\it{L}_{\rm{line}}}{10^{44}erg~s^{-1}}\right)  
	+ \rm{2~log}\left(\frac{\rm{FWHM}}{\rm{km~s^{-1}} }\right) ,
\end{equation}
where $c = 1.70 \pm 0.07$ and $d = 0.63 \pm 0.00$, also determined by SH11.

	For the low-$z$ blazar SDSS J083353.88+422401.8, Mg~II is not in the observed spectral
wavelength range, and so the Eddington ratio must be calculated using other BELs.
H$\alpha$ is the only measurable BEL in this blazar, as H$\beta$ is too weak to be discernible 
from the continuum, even in the low continuum luminosity epochs. We thus estimate 
$M_{\rm{BH}}$ using its relation with the H$\alpha$ line luminosity and FWHM, determined by 
SH11 (their Equation 10) and based on \citet{gr05}. Because SH11 did not measure correlations 
between continuum and line luminosities for H$\alpha$, we estimate the continuum luminosity 
$L_{\rm{5100}}$ by first estimating the FWHM of H$\beta$ (not observable) using its correlation 
with the H$\alpha$ FWHM determined in SH11. We can then estimate $L_{\rm{5100}}$ from its 
relation with the $M_{\rm{BH}}$ and the H$\beta$ FWHM, using Equation 5 from \citet{sh12} with 
their constants $a = 0.672$ and $b = 0.61$, determined by \citet{mc04}. Finally, the bolometric luminosity 
is calculated using $L_{\rm{bol}}$/$L_{\rm{5100}}$ = 9.26 \citep{ri06}.

	We estimate the $M_{\rm{BH}}$ and $L_{\rm{bol}}$/$L_{\rm{Edd}}$ of each blazar using
the spectral epochs for which the BELs have the largest EWs; these values are also listed in Table 1.
All uncertainties on these derived quantities include the propagated statistical uncertainties from the 
spectral line fitting, as well as the reported statistical and systematic uncertainties of the constants in 
each of the scaling relations. We are also able to estimate $L_{\rm{bol}}$ and $M_{\rm{BH}}$ for 
SDSS J101603.13+051302.3 using C~IV independently in addition to Mg~II; this calculation is performed using
$a = 0.863 \pm 0.009$ and $b = 7.66 \pm 0.41$ in Equation 1 (SH11), where the line luminosity is 
for C~IV and the continuum luminosity is for 1350\AA, and using a bolometric correction of 
$L_{\rm{bol}}$/$L_{\rm{1350}}$ = 3.81 \citep{ri06}. The $M_{\rm{BH}}$ based on C~IV is then 
calculated by setting $c = 1.52 \pm 0.22$ and $d = 0.46 \pm 0.01$ in Equation 2 (SH11). The 
$M_{\rm{BH}}$ and $L_{\rm{bol}}$/$L_{\rm{Edd}}$ are calculated independently from C~IV and Mg~II in 
this object are consistent to within the uncertainty. 

	We also estimate the total luminosity of the broad line emission to estimate the Eddington ratio
of the BELs ($L_{\rm{BLR}}$/$L_{\rm{Edd}}$). This calculation is done similar to the method of \citet{ce97},
who found that the ratio of the total luminosity of all BELs to the line luminosities ($L_{\rm{BLR}}$/$L_{\rm{line}}$) for 
H$\alpha$, Mg~II, and C~IV are 555.76/77, 555.76/34, 555.76/63, respectively, based primarily on BEL flux 
ratios in the quasar composite spectrum of \citet{fr91}. Using these factors, 
we calculate $L_{\rm{BLR}}$ for each of our transition blazars at the spectral epoch with largest EW, 
similar to that done for $L_{\rm{bol}}$/$L_{\rm{Edd}}$ above. For SDSS J101603.13+051302.3, 
the $L_{\rm{BLR}}$ independently based on C~IV is also highly consistent with Mg~II. Although a single 
measurement of $L_{\rm{bol}}$/$L_{\rm{Edd}}$ is most helpful for classification in cases where multiple BEL 
are available (e.g. by averaging estimates derived from multiple lines), we opt to use the estimate from Mg~II 
as the fiducial $L_{\rm{bol}}$/$L_{\rm{Edd}}$ for J101603.13+051302.3 due to its smaller scatter in the 
$M_{\rm{BH}}$ scaling relations we use in comparison to that of C~IV, which instead serves as a helpful 
consistency check. Figure 10 presents 
the bolometric Eddington ratio of the transition blazars against the Eddington ratio of the BELs for the 
transition blazars in our sample, as well as for the blazar which did not transition. The Eddington ratios of 
these blazars are similar to the Eddington ratios of typical FSRQs (and much larger than those from 
radiatively inefficient accretion flows in BLLs); we discuss our interpretation of these results in 
Section 5.

\subsection{Spectral Energy Distribution Properties}

	Blazars are well-known to emit strongly at a wide range of wavelengths, and their SEDs are 
dominated by synchrotron emission at radio up to soft X-ray wavelengths, and inverse-Compton 
emission at hard X-ray to $\gamma$-ray wavelengths. The peaks of the synchrotron and 
inverse-Compton emission tend to occur at lower frequencies for FSRQs than for BLLs in general, 
although BLLs have often been further subdivided into low-, intermediate-, and high-synchrotron 
peaked BLLs, leading to the idea of a possible continuous `blazar sequence' of SEDs \citep{fo98, gh08}. 
We compile SEDs for the transition blazars in our sample using the ASI Science Data Center Virtual 
Observatory SED builder\footnote{http://tools.asdc.asi.it/SED/} \citep{st11}, conveniently linked to all 
blazars in the ROMA-BZCAT catalog. The archival multi-wavelength data available are most complete 
at radio to optical wavelengths, and visual inspection of the SEDs of our sample of transition blazars 
shows that the synchrotron peak is well-determined for all but one (SDSS J143758.67+300207.1), 
which has few data points in its SED. 

	We estimate the frequency and luminosity of the synchrotron peak of the transition blazars 
(except for SDSS J143758.67+300207.1) by fitting a third-order polynomial to the synchrotron emission 
in the SEDs. In the fitting, we have purposely excluded data points at frequencies $>$10$^{17}$ Hz 
to avoid inverse-Compton emission. The estimated synchrotron peak frequencies and luminosities
are listed in Table 2. These order of magnitude estimates are approximate since we are not fitting a 
physically-motivated model to the SED, and the archival SED data points are not contemporaneous. 
However, the SED peaks are reasonably well-determined, and sufficient for classification of SED type. 
All the measured peak frequencies of these transition blazars are similar to the $10^{13-14}$Hz range of 
low-synchrotron peaked blazars, and their luminosities are also consistent with the 
$>$10$^{43}$ erg s$^{-1}$ luminosities of blazars in this SED class. This result is in strong contrast to the 
$\sim10^{14-16}$ Hz and $\sim10^{16-18}$ Hz synchrotron peak frequencies of intermediate- and 
high-synchrotron peaked blazars, respectively \citep {ab10a}. Thus, the SED properties of these transition blazars (as well as SDSS J083223.22+491321.0) are consistent with FSRQs and low-synchrotron 
peaked BLLs.

\begin{figure}
\begin{center}
\includegraphics[width=0.48\textwidth]{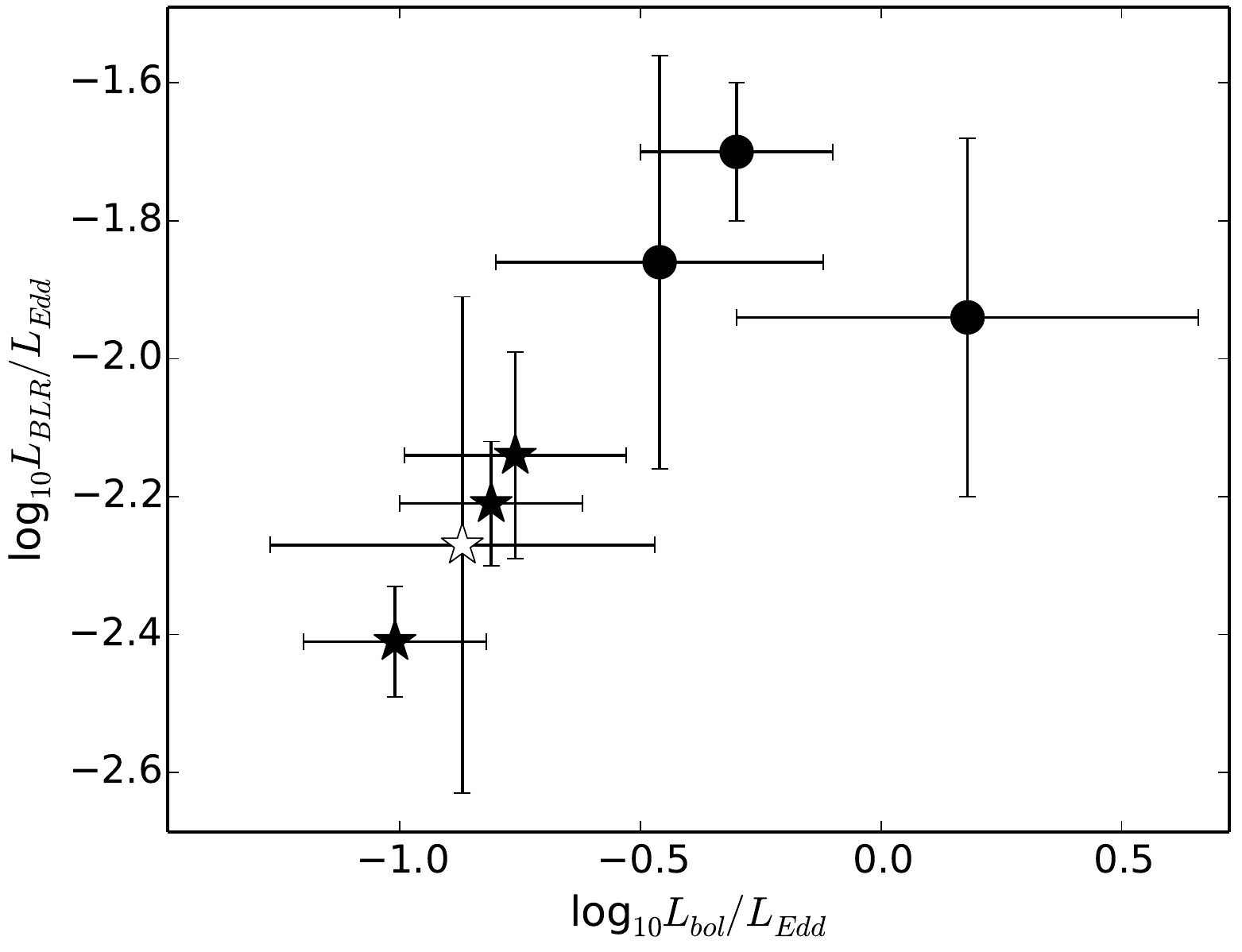}
\caption{The bolometric Eddington ratio versus the Eddington ratio of the total broad emission line luminosity
for the six transition blazars in our sample (filled symbols), and the additional blazar which did not transition (open star). These values are listed in Table 1. Blazars with \emph{Fermi} detections are shown with stars, and without are shown 
with dots. These Eddington ratios are much larger than the $L_{\rm{bol}}$/$L_{\rm{Edd}}$ $\sim$ 0.01 typical of
radiatively inefficient accretion flows in BLLs, and are instead consistent with those of classical thin accretion disks in FSRQs.}
\end{center}
\end{figure}

\begin{deluxetable*}{ccccccccccc}
\tablecolumns{12}
\tablewidth{0pt}
\tablecaption{$\gamma$-ray and long-term photometric variability properties of the transition blazars in our sample.}
\tablehead {\colhead{SDSS} & \colhead{log$_{10}$$L_{\gamma}$$^\textrm{a}$ } & \colhead{$\alpha_{\gamma}$$^\textrm{b}$ } & \colhead{log$_{10}\nu_{peak}^\textrm{c}$} & \colhead{log$_{10}L_{peak}^\textrm{d}$} & \colhead{log$_{10}$SF$_\infty$$^\textrm{e}$} & \colhead{log$_{10}$$\tau$$^\textrm{f}$}
\\
\colhead{Object} & \colhead{[erg s$^{-1}$]} & & \colhead{[Hz]}  & \colhead{[erg s$^{-1}$]} & \colhead{[mag]} & \colhead{[days]}} 
\startdata
J083223.22+491321.0 & & & 13.12 & 45.51 & $-$0.14 & 1.33 \\

J083353.88+422401.8 & 45.22 & 2.33 & 14.11 & 44.80 & $-$0.21 & 1.38  \\

J101603.13+051302.3 & 47.99 & 2.08 & 13.27 & 45.94 & 0.31 & 0.79 \\

J125032.57+021632.1 & & & 13.74 & 45.49 & 0.05 & 1.76  \\

J130823.70+354637.0 & 46.89 & 2.28 & 12.88 & 45.89 & $-$0.09 & 0.81\\

J143758.67+300207.1 & & & & &  0.13 & 0.93 \\

J220643.28$-$003102.5 & 46.61 & 2.16  & 13.04 & 46.02 & $-$0.06 & 0.78  

\enddata
\tablenotetext{a}{0.1 - 100 GeV luminosity from Nolan et al. (2012)}
\tablenotetext{b}{0.1 - 100 GeV photon number power-law index from Nolan et al. (2012)}
\tablenotetext{c}{Estimated frequency of the synchrotron peak in the SED.}
\tablenotetext{d}{Estimated luminosity of the synchrotron peak in the SED.}
\tablenotetext{e}{Optical photometetric variability amplitude on long timescales.}
\tablenotetext{f}{Characteristic timescale of optical photometric variability in the rest-frame.}
\end{deluxetable*}

\subsection{$\gamma$-ray Properties}

	The Large Area Telescope \citep{at09} aboard \emph{Fermi}'s has surveyed the $\gamma$-ray sky to unprecedented 
depth, thus far detecting 1,873 $\gamma$-ray sources in the 2-year 2FGL catalog \citep{no12}, 
including 1,121 AGN in the 2nd \emph{Fermi} AGN catalog \citep{ac11}. The overwhelming majority 
of \emph{Fermi}-detected AGN are blazars, approximately evenly divided between FSRQs and 
BLLs. Analysis of these \emph{Fermi} blazars has revealed that BLLs tend to have $\gamma$-ray 
luminosities $L_\gamma$ $\lesssim$ 10$^{46}$ erg s$^{-1}$ and $\gamma$-ray spectral indices 
$\alpha_{\gamma}$ $\lesssim$ 2.2 in the 0.1 - 100 GeV band, while FSRQs exceed these values. 
However, there is significant overlap in these $\gamma$-ray properties between the two subclasses 
\citep{ac11}, likely due to some combination of large measurement uncertainties and misclassification, 
in addition to astrophysical origins.

	We match our transition blazars to the \emph{Fermi} 2FGL catalog to investigate their $\gamma$-ray
properties. Four of the blazars in our sample are matched to 2FGL sources that have association probabilities
$>$95\% (i.e. the blazar is highly likely to be the optical counterpart of the $\gamma$-ray source); 
their $\gamma$-ray luminosities and spectral indices are also listed in Table 2. None of the 
four transition blazars in our sample detected by \emph{Fermi} can be clearly classified based on their 
$\gamma$-ray properties, and all lie in the area occupied by FSRQs and low-to-intermediate 
synchrotron peaked BLLs. \citet{gh11} argued that for such objects, a better classifier to use is
the Eddington ratio based on the BEL ($L_{\rm{BLR}}$/$L_{\rm{Edd}}$), with a dividing value of 
$\sim$5$\times$10$^{-4}$. We note that although synchrotron peak frequencies and luminosities 
of these transition blazars (where determined) are similar, not all are detected in $\gamma$-rays. 

\subsection{Optical Variability Properties}

	Blazars are well-known to be among the most variable extragalactic objects detected in 
large-scale optical time-domain surveys \citep{ba09}. An investigation of long-term optical light curves 
recalibrated by \citet{se11} from the Lincoln Near-Earth Asteroid Research \citep[LINEAR,][]{sto00} survey 
by \citet{ru12} revealed that the jet synchrotron-dominated continuum emission in blazars is systematically 
more variable than the thermal disk-dominated 
continuum in typical Type 1 quasars. In particular, \citet{ru12} found that blazars have typical 
optical variability amplitudes of $\sim$0.5 mag in the $r$-band on rest-frame timescales greater than 
$\sim$100 days; below this characteristic timescale, the optical variability is self-correlated 
and smaller in amplitude. The strong variability amplitudes and short variability timescales 
observed in comparison to typical quasars was interpreted by \citet{ru12} as due to the effects of 
Doppler boosting in the relativistic jet of blazars.

	We compile publicly-available optical light curves of the transition blazars in our sample from 
the LINEAR and Catalina Sky Survey/Catalina Real-Time Transient Survey \citep{dr09} databases
to investigate their optical variability properties. The optical light curves constructed using these 
two surveys are also shown in Figures 2 to 8; they are well-sampled, with 98 - 978 epochs (median 
of 444 epochs) spread over $\sim$11 years in the observed-frame. All transition blazars in our 
sample display strong optical variability, with maximum amplitude on the range of $\sim$1.5 to 6 
mags in the data. 

	To compare the optical variability properties of these transition blazars to the large sample of
FSRQs and BLLs from \citet{ru12}, we fit each of these light curves to a first-order continuous 
autoregressive model (i.e., a damped random walk) using the method of \citet{ko10}, and discussed
in detail in \citet{ma10}. Figure 11 compares the rest-frame characteristic timescale of variability $\tau$,
and the variability amplitude on long timescales (when the variability is uncorrelated and strongest)
SF$_\infty$, of these seven blazars to the large sample of 60 bright blazars from \citet{ru12}. The seven 
blazars in our sample have larger variability amplitudes and shorter timescales of variability 
than typical blazars. A two-sample Kolmogorov-Smirnov test between the distributions of $\tau$ 
and SF$_\infty$ in Figure 11 for our transition blazars and the blazar sample of \citet{ru12} results
in a p-value of $3.1\times10^{-3}$ for $\tau$, and $5.6\times10^{-4}$ for SF$_\infty$.

	Although the light curves we construct use data from both LINEAR and the 
Catalina surveys, the filters and calibrations are not exact between these two surveys; this issue will lead to 
a systematic offset between their measured magnitudes. However, both surveys used white-light filters 
and the systematic offset in magnitude is small ($<$0.1 mag), especially in comparison to the 
several-magnitude variability observed in our transition blazar light curves. Thus, any differences in 
calibration between the two surveys are dwarfed by the intrinsic variability of transition blazars,
and do not significantly affect our analysis. 

\begin{figure}
\begin{center}
\includegraphics[width=0.49\textwidth]{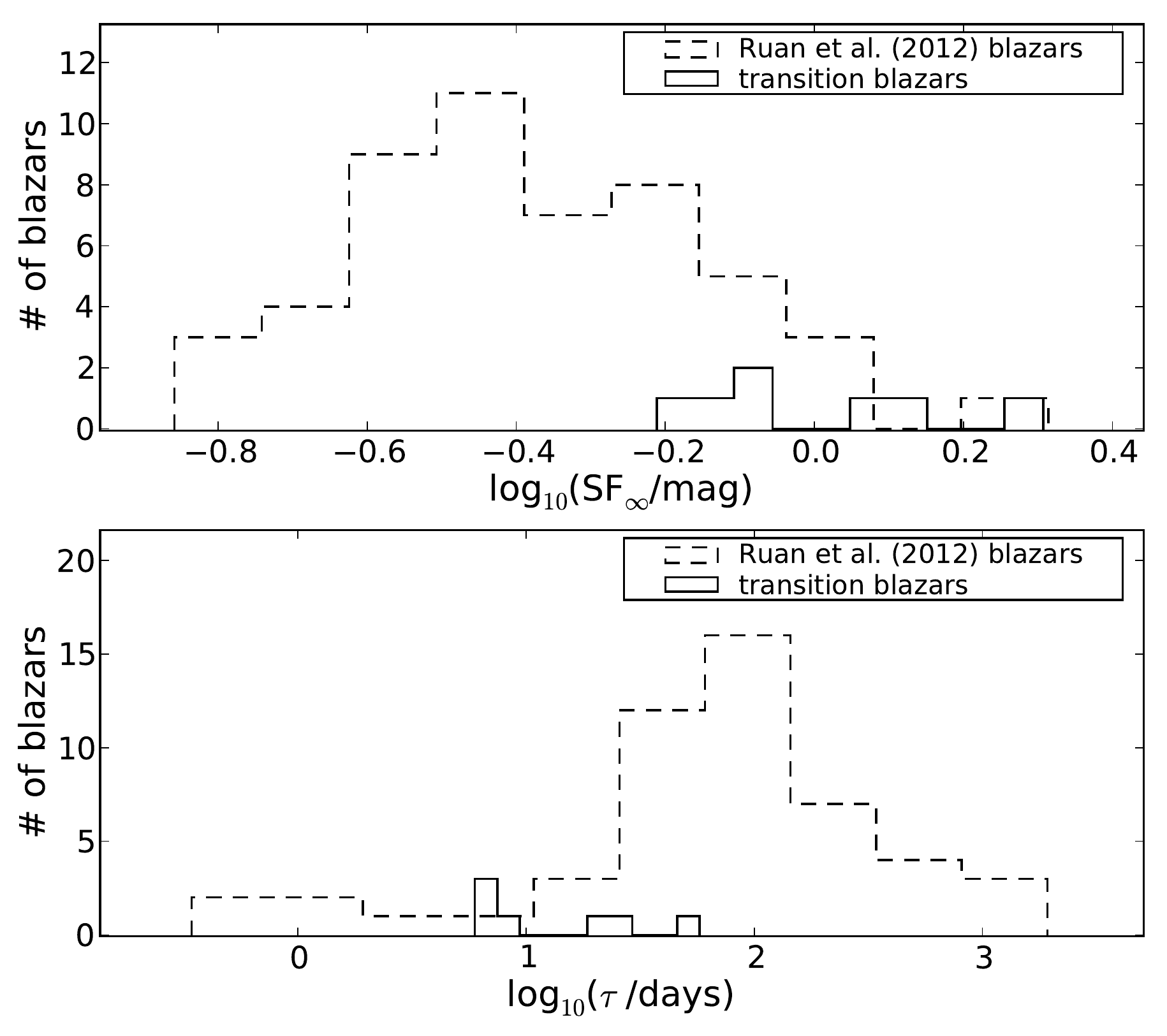} 
\caption{Top: Histogram of the variability amplitude on long timescales for the long-term photometric
light curves of the seven blazars in our sample modeled as a damped random-walk, as well as for a sample 
of 60 blazars from \citet{ru12}. Bottom: Similar to the top panel, but for the characteristic timescale 
of variability in the rest-frame. The seven blazars in our sample have significantly larger variability 
amplitudes and shorter timescales of variability than typical blazars, supporting the interpretation that the 
jet emission is particularly strongly-beamed.}
\end{center}
\end{figure}

\section{Notes on Individual Objects}

\emph{SDSS J083223.22+491321.0}: The Mg~II line EW of this blazar is $<$5\AA~in all three available 
epochs of spectra (Figure 8), and thus this blazar appears to be a canonical BLL and was 
included in our analysis for completeness due to its newly-determined 
redshift of $z = 0.548$. However, despite its small Mg~II EW and under-luminous Mg~II emission 
relative to the FSRQs in \citet{sh12}, its high bolometric Eddington ratio instead strongly suggests 
that it is in fact a FSRQ, and would transition with further spectroscopic monitoring. This blazar is not 
\emph{Fermi}-detected.

\emph{SDSS J083353.88+422401.8}: The H$\alpha$ line is visible in this low-redshift blazar in all three 
available epochs of spectra (Figure 2), although its EW is small in even the largest-EW epoch. 
\citet{gi12} suggested that blazars with BELs that transition in their optical spectra likely have highly 
luminous H$\alpha$ emission in the observed-frame infrared with EWs that remain $>$5\AA. However, 
this blazar is an example in which even the H$\alpha$ emission transitions. The $M_{\rm{BH}}$ and 
H$\alpha$ luminosity we calculate for this blazar are smaller than typical of FSRQs in the sample of 
\citet{sh12}, though this may be affected by our indirect method of estimating the continuum luminosity 
through correlations between the H$\alpha$ and the unobserved H$\beta$ BELs (see Section 3.1). 
To verify our $M_{\rm{BH}}$ estimate, we also fit the H$\alpha$ FWHM in the second-largest EW spectral 
epoch (MJD = 54524) for this blazar, and compare to the largest EW epoch. The FWHMs of both 
epochs are highly consistent, and thus our relatively small $M_{\rm{BH}}$ estimate appears to 
be reliable. This blazar is \emph{Fermi}-detected, and its optical light curve is exceptionally 
well-sampled, showing variability up to $\sim$2 magnitudes.

\emph{SDSS J101603.13+051302.3}: The C~IV, C~III], and Mg~II BELs all show strong 
variability in EW (up to a factor of $>$60) over five available epochs of SDSS spectra (Figure 3). 
The decrease in continuum luminosity in these spectra causes the BELs to transition nearly in 
unison. This strong EW variability is not particularly surprising, as the long-term continuum variability 
probed by the the broadband optical light curve varies up to $\gtrsim$5 magnitudes. This blazar is 
a bright \emph{Fermi} source, and its bolometric Eddington ratio determined from Mg~II (more robust 
than that from C~IV) shows it is accreting even more rapidly than typical FSRQs.

\emph{SDSS J125032.57+021632.1}: Although this blazar did not previously have a redshift 
in the ROMA-BZCAT catalog, the Mg~II line is well-detected in both epochs of spectra (Figure 4), 
and we determine a new robust redshift of $z = 0.959$. This blazar is not \emph{Fermi} detected, 
and has an Eddington ratio of $\sim$0.1, more similar to typical quasars than FSRQs.

\emph{SDSS J130823.70+354637.0}: The two epochs of spectra for this blazar are 
observed during dramatic dimming and flaring events in its optical light curve (Figure 5). 
These dramatic flares of $\sim$2 magnitudes appear to be a frequent occurrence for this 
blazar, on timescales as short as days to weeks. The Mg~II line is well-detected in both epochs, 
and this blazar is \emph{Fermi}-detected.

\emph{SDSS J143758.67+300207.1}: The Mg~II line in this blazar is weak in both epochs 
of spectra (Figure 6), but the strong continuum variability causes its EW to increase by a factor 
of $>$60. The redshift for this blazar ($z = 1.230$) is new, and this blazar is not \emph{Fermi} 
detected. Although the optical light curve is sparsely sampled, large variations of up to $\sim$3 
magnitudes are evident. 

\emph{SDSS J220643.28-003102.5}: This \emph{Fermi}-detected blazar's Mg~II line is much 
more clearly detected in the fainter-continuum epoch (Figure 7). The optical light curve shows 
strong variability, with large ($\sim$2 magnitude) flares on timescales of days to weeks. 
Its characteristic timescale of variability of 6 days in the rest-frame is the shortest among the transition 
blazars in our sample. This blazar is in the Stripe 82 multi-epoch imaging portion of SDSS, and its 
SDSS optical light curve is among the most variable of all quasars in \citet{ma10}. 

\section{Discussion}
	In the previous sections, we have explored the spectral, temporal, and multi-wavelength 
properties of our sample of transition blazars; in this section, we will piece together these properties 
to understand the nature of these objects. The Eddington ratios of the transition blazars in our sample 
calculated in Section 3.1 (and shown in Figure 10) are more than an order of magnitude above the 
approximate $L_{\rm{bol}}$/$L_{\rm{Edd}}$ $\sim$ $0.01$ value that typically separates the radiatively 
efficient and inefficient accretion regimes in AGN, and are even higher than the $L_{\rm{bol}}$/$L_{\rm{Edd}}$
$\sim$ $0.1$ typical of Type 1 quasars in SDSS (SH11). This result strongly suggests that  the transition 
blazars in our sample are actually rapidly-accreting FSRQs, which have been found to have 
systematically large bolometric Eddington ratios of $L_{\rm{bol}}$/$L_{\rm{Edd}}$ $\sim$ 0.1 to 1 in 
a large \emph{Fermi}-selected sample \citep{sh12}. The $L_{\rm{BLR}}$/$L_{\rm{Edd}}$ of these 
transition blazars are correspondingly also well above the $\sim$$5\times10^{-4}$ divide suggested by 
\citet{gh11} to separate FSRQs from BLLs. We emphasize that our estimates of $L_{\rm{bol}}$/$L_{\rm{Edd}}$
for each transition blazar are not strongly affected by contamination from jet emission since the continuum
luminosities (without jet contamination) are not directly measured in the spectrum, but rather indirectly estimated
based on scaling relations with the BEL luminosities as discussed in Section 3.1.

	Our FSRQ-like interpretation of these blazars is further supported by the strong variability 
observed in the BEL luminosities of up to $\sim$0.5 dex (Figure 9, left panel). This effect is likely 
due to the strong flux variability of the ionizing thermal continuum from radiatively-efficient thin accretion 
disks that are thought to exist in FSRQs, in contrast to the radiatively-inefficient accretion flows in BLLs,
since variations in the jet continuum only weakly correlates with BEL flux variability \citep{is13}.
The FSRQ nature of these transition blazars is also supported by our investigation of their SEDs 
(Section 3.2), which revealed that their synchrotron peaks are at frequencies of $\sim$10$^{12-13}$ Hz, 
and luminosities of $10^{45-46}$ erg s$^{-1}$; these SED properties are similar to FSRQs and low 
synchrotron-peaked BLLs, although the underlying accretion state is difficult to discern from the SED alone.

	Although swamping of BELs by the jet continuum in transition blazars as suggested by the observed 
anti-correlation between the line EWs and continuum luminosities (see Figure 9) may be expected
in FSRQs, the results of our systematic search reveals that such cases are instead uncommon. Our 
investigation suggests that the transition phenomenon occurs in rare FSRQs with particularly strongly-beamed 
jets, oriented at extremely small angles along the line of sight. The particularly strong beaming in such objects 
would lead to correspondingly strong jet dominance of the optical spectra, as well as strong flux variability. 
This interpretation is supported by our investigation of the long-term optical variability of these transition 
blazars in Section 3.4: they show larger amplitudes and shorter timescales of variability (Figure 11) 
than typical blazars. This uniquely strong variability in transition blazars suggests that the jet beaming is 
especially strong in these objects, with large Lorentz factors. 

	The transitional nature of the BEL EWs in these blazars may lead to misclassifications if the subclass
classification is based solely on EWs in single-epoch spectra as traditionally done. Although transition
blazars are rare, misclassification can still significantly affect investigations of blazar redshift evolution 
at high redshifts due to their rarity. Our findings suggest that some BL Lacs may actually be strongly-beamed 
FSRQs, thus decreasing the apparent BL Lac population, especially at the higher redshifts of FSRQs and 
partially due to the high luminosities of these particularly strongly-beamed blazars. This effect can decrease the 
small sample sizes of high-redshift BL Lacs used in investigations of their redshift evolution, and may even add
credence to claims of a BL Lac negative redshift evolution. Since the SEDs of the transition blazars in our 
sample are ambiguous between FSRQs and low synchrotron-peaked BLLs despite their unambiguously large 
accretion rates, classification based on SEDs can also lead to incorrect results. Even when multi-epoch spectra are 
available (thus allowing for identification of transition blazars), some studies have incorrectly assumed 
that transitional blazars in which all BELs have EW $<$ 5\AA~in \emph{any} epoch are BLLs 
\citep[e.g.,][]{sh12, sh13a}. In contrast, our investigation reveals that these transition blazars are 
instead likely to be FSRQs. 

	Similar to the arguments presented in \citet{gh11}, we strongly recommend that 
classification of ambiguous blazars be based on Eddington ratios, especially through the use of 
multi-epoch spectra so that the BEL properties can be robustly determined from the large-EW spectral 
epochs. This Eddington ratio-based classification is useful for not only transition blazars, but also for 
blazars such as SDSS 083223.22+491321.0 (Figure 8), which does not have BEL EW $>$ 5\AA~
in any of the observed epochs. Although this blazar appears to be a canonical BLL, its Eddington 
ratio of $L_{\rm{bol}}$/$L_{\rm{Edd}}$ $\sim$ $0.17$ instead clearly reveals it to be a 
FSRQ with extreme swamping of the BELs. Additionally, the large EW variations in transition blazars 
observed in multi-epoch spectroscopy also aids robust redshift determinations.

	Finally, despite the strong beaming in their jets that cause strong photometric variability
and EW changes in the BELs, it is puzzling that not all of these transition blazars are $\gamma$-ray 
detected. The synchrotron peak frequencies and luminosities in SEDs of the transition blazars in 
our sample are all in a similar range, without any clear distinction between the $\gamma$-ray loud
and quiet blazars. It is possible that the $\gamma$-ray luminosity in these jets with extreme beaming
may be suppressed due to strong $\gamma\gamma$ attenuation from pair production \citep{be95,bl95} 
or Klein-Nishina effects in the electron Compton cooling \citep[e.g.][]{mo05}. Many of the transition 
blazars in our sample have detections in \emph{GALEX}, \emph{Swift}, and \emph{ROSAT} in addition to \emph{Fermi}; future investigations of the high-energy inverse-Compton SED peaks of these blazars will 
be able to shed light on this problem. Future studies of $\gamma$-ray variability in these transition 
blazars will also be able probe the physical mechanism of transitions between jet-dominated
and disk-dominated states \citep{ch13}.

\section{Conclusions}

Using a large sample of 602 pairs of repeat blazar spectra of 354 blazars from SDSS, we 
present a systematic search for blazars that have transitioned between the BLL and FSRQ 
subclasses based on the EW of the BELs. These repeat spectral observations span timescales 
from intra-night to $\sim$12.5 years in the observed-frame, and are selected by a variety of 
multi-wavelength properties. We find six clear cases of transition blazars which show dramatic 
variability in their BEL EWs; the strong BEL EW variability allows us to determine new redshifts 
for two of the transition blazars in our sample, as well as one additional blazar. We investigate
the SDSS spectral, $\gamma$-ray, SED, and optical variability properties of these transition 
blazars to understand their nature as probed by these methods. Our key results
are as follows: 

1) EWs of BELs in transition blazars can vary dramatically, with observed variations up to factors of 
$>$60. Transition blazars have high accretion rates, with bolometric Eddington 
ratios typical of FSRQs. A clear anti-correlation between the continuum luminosity and BEL
EWs strongly suggests that the transitional phenomenon is primarily due to swamping of the BELs
by the highly variable jet continuum. However, the luminosities of the BELs are also 
variable, further supporting the existence of radiatively efficient thin accretion disks in these objects.

2) The swamping of the BELs in transition blazars occurs when the jet continuum is strongly-beamed, 
placing transition blazars as rare cases of FSRQs with particularly strongly-beamed jets. 
This interpretation is supported by analysis of their long-term photometric light 
curves, which display larger amplitudes and shorter timescales of variability than typical 
blazars, consistent with the effects of strongly-beamed jet continuum emission. 

3) The SEDs of these transition blazars show low-frequency, luminous synchrotron 
peaks, similar to FSRQs and low-peaked BLLs. Their $\gamma$-ray properties (where detected) are 
also ambiguous between these two classes. Since classification based on BEL EWs, SEDs, and 
$\gamma$-ray properties of these objects is difficult, we suggest the use of Eddington ratios. 
 
Although transition blazars are rare, correct classification will impact studies of the divergent 
properties of BLLs and FSRQs, especially regarding their redshift evolution. Future directions for 
investigation of these rare transition blazars can focus on understanding the nature of the strong 
beaming using radio observations, and study of their high-energy inverse-Compton SED peaks to 
understand why only some of these strongly-beamed FSRQs are $\gamma$-ray loud. Future surveys 
such as the Time-Domain Spectroscopic Survey \citep{gr14} of SDSS-IV will include more repeat spectra of 
blazars, and will uncover many more such transitional cases.

\acknowledgments
JJR thanks James R.~A.~Davenport for helpful discussions. Support for JJR was provided by 
NASA through Chandra Award Numbers AR9-0015X, AR0-11014X, and AR2-13007X, 
issued by the Chandra X-ray Observatory Center, which is operated by the Smithsonian Astrophysical 
Observatory for and on behalf of NASA under contract NAS8-03060. Support for WNB was 
provided by NSF grant AST-1108604 and NASA ADP grant NNX10AC99G.

Funding for SDSS-III has been provided by the Alfred P. Sloan Foundation, the Participating Institutions, the National Science Foundation, and the U.S. Department of Energy Office of Science. The SDSS-III web site is http://www.sdss3.org/.

SDSS-III is managed by the Astrophysical Research Consortium for the Participating Institutions of the SDSS-III Collaboration including the University of Arizona, the Brazilian Participation Group, Brookhaven National Laboratory, Carnegie Mellon University, University of Florida, the French Participation Group, the German Participation Group, Harvard University, the Instituto de Astrofisica de Canarias, the Michigan State/Notre Dame/JINA Participation Group, Johns Hopkins University, Lawrence Berkeley National Laboratory, Max Planck Institute for Astrophysics, Max Planck Institute for Extraterrestrial Physics, New Mexico State University, New York University, Ohio State University, Pennsylvania State University, University of Portsmouth, Princeton University, the Spanish Participation Group, University of Tokyo, University of Utah, Vanderbilt University, University of Virginia, University of Washington, and Yale University.

The LINEAR program is sponsored by the National Aeronautics and Space 
Administration (NRA No. NNH09ZDA001N, 09-NEOO09-0010) and the United States 
Air Force under Air Force Contract FA8721-05-C-0002. Opinions, interpretations, conclusions, 
and recommendations are those of the authors and are not necessarily endorsed by the 
United States Government.

The Catalina Sky Survey is funded by the National Aeronautics and Space
Administration under Grant No. NNG05GF22G issued through the Science
Mission Directorate Near-Earth Objects Observations Program.  The CRTS
survey is supported by the U.S.~National Science Foundation under
grants AST-0909182.

Part of this work is based on archival data, software or on-line services provided by the ASI Science Data Center (ASDC). 

\bibliography{bibref}
\bibliographystyle{apj}

\end{document}